%% file: main.tex
\documentclass[preprints,article,accept,pdftex,moreauthors]{Definitions/mdpi}

\newcommand{\bl}[1]{{#1}}

\usepackage{amsmath,amsfonts,amssymb,bm}
\usepackage{algorithm}
\usepackage{array}
\usepackage{booktabs}
\usepackage{multirow}
\usepackage{siunitx}

\firstpage{1}
\pubvolume{1}
\issuenum{1}
\articlenumber{0}
\pubyear{2024}
\copyrightyear{2024}
\datereceived{ }
\daterevised{ } 
\dateaccepted{ }
\datepublished{ }
\hreflink{https://doi.org/} 

\usetikzlibrary{arrows,backgrounds,calc,positioning,shapes,shadows}
\usetikzlibrary{arrows.meta}
\usetikzlibrary{decorations.pathreplacing}
\usetikzlibrary{calligraphy}
\usetikzlibrary{shapes.multipart, shapes.misc}
\usetikzlibrary{plotmarks}

\definecolor{color0}{rgb}{1,0.4980392157,0.05490196078} 
\definecolor{color1}{rgb}{0.1215686275,0.4666666667,0.7058823529} 
\definecolor{color2}{rgb}{0.1725490196,0.6274509804,0.1725490196} 
\definecolor{color3}{rgb}{0.8392156863,0.1529411765,0.1568627451} 
\definecolor{color4}{rgb}{0.5803921569,0.4039215686,0.7411764706} 

\usepackage{pgfplots}
  \pgfplotsset{compat=newest}
  \usetikzlibrary{plotmarks}
  \usepgfplotslibrary{patchplots}
  \usepackage{grffile}
\usepgfplotslibrary{groupplots}
\pgfplotsset{every axis/.append style={
                    label style={font=\footnotesize},
                    tick label style={font=\footnotesize},
                }}

\Title{Trustworthiness for an Ultra-Wideband Localization Service}

\TitleCitation{Trustworthiness for an Ultra-Wideband Localization Service}

\Author{Philipp Peterseil $^{1\orcidA{}}$, Bernhard Etzlinger $^{1}\orcidB{}$, Jan Hor\'{a}\v{c}ek $^{1, 3}$, Roya Khanzadeh $^{1,2\orcidD{}}$ and Andreas Springer $^{1\orcidE{}}$}

\AuthorNames{Philipp Peterseil, Bernhard Etzlinger, Jan Hor\'{a}\v{c}ek, Roya Khanzadeh, Andreas Springer}

\AuthorCitation{Peterseil, P.; Etzlinger, B.; Hor\'{a}\v{c}ek, J.; Khanzadeh, R.;  Springer, A.}

\address{%
$^{1}$ \quad Johannes Kepler University Linz; \{firstname.lastname\}@jku.at\\
$^{2}$ \quad JKU SAL IWS Lab \\
$^{3}$ \quad LIT Secure and Correct Systems Lab}

\abstract{
Trustworthiness assessment is an essential step to
assure that interdependent systems perform critical functions as anticipated, even under adverse conditions. In this paper, a holistic trustworthiness assessment framework for ultra-wideband self-localization is proposed, including attributes of reliability, security, privacy, and resilience.
Our goal is to provide guidance for evaluating a system’s trustworthiness based on objective evidence, so-called trustworthiness indicators. These indicators are carefully selected through the threat analysis of the particular system. Our approach guarantees that the resulting trustworthiness indicators correspond to chosen real-world threats. Moreover, experimental evaluations are conducted to demonstrate the effectiveness of the proposed method.  While the framework is tailored for this specific use case, the process itself serves as a versatile template, which can be used in other applications in the domains of the Internet of Things or cyber-physical systems.}

\keyword{Trustworthiness; Ultra Wideband; Indoor Localization}

\begin{document}
\crefname{chapter}{Chap.}{Chaps.}
\crefname{section}{Sec.}{Secs.}
\crefname{subsection}{Sec.}{Secs.}
\crefname{figure}{Fig.}{Figs.}
\crefname{table}{Tab.}{Tabs.}
\crefname{equation}{Eq.}{Eqs.}
\crefname{appendix}{App.}{Apps.}
\crefname{theorem}{Thm.}{Thms.}
\crefname{lemma}{Lem.}{Lems.}
\crefname{algorithm}{Alg.}{Algs.}



\section{Introduction}

Whenever systems interact with each other and with the physical world, preserving integrity to perform mission-critical tasks is essential. Therefore, in computer security, the concept of trust ensures that each component of software and hardware can be relied upon \cite{feng2017trusted}. The notion of trust was adopted by the United States National Institute of Standards and Technology, which defined trustworthiness as one of nine essential aspects of cyber-physical systems \cite{NIST2016cpsframework}. The term has also been adopted by the Internet of Things (IoT) community \cite{buchheit30industrial} and has even been leveraged as a key value indicator for the International Mobile Telecommunications 2030 vision for the future 6G communication standard.
While these developments highlight the importance of trustworthiness, it remains a vague term in the literature. Beyond a unified understanding of trustworthiness, practical frameworks that link high-level definitions to concrete realizations are lacking.

\bl{Ultra-wideband (UWB) localization services provide accurate positioning. The high bandwidth of typically \SI{500}{\mega\hertz} allows precise distance measurements based on the time-of-flight of the radio signals. This technology is particularly advantageous for indoor environments where traditional GPS is ineffective. UWB localization offers centimeter-level accuracy, making it suitable for applications such as asset tracking and indoor navigation.}

To address the gap in trustworthiness assessment, a systematic method is proposed for the IoT use case of UWB \textit{self-localization}, i.e., specifically for a single node estimating its position relative to multiple anchors. Throughout the paper, the following definitions are used:
\begin{itemize}
    \item \textit{Trustworthiness:} "Trustworthiness is the demonstrable likelihood that the system performs according to designed behavior under any set of conditions as evidenced by characteristics including, but not limited to, safety, security, privacy, reliability and resilience." \cite{NIST2016cpsframework}
    \item \textit{Trustworthiness metric:}
    A trustworthiness metric is considered to be any measurement instance that describes the trustworthiness level of system operation.
    \item \textit{Trustworthiness indicator:} 
    Trustworthiness indicators map trustworthiness metrics to a likelihood interval in the range $[0,1]$, where 0 represents the lowest level of trustworthiness and 1 represents the highest level.
    \item \textit{Trustworthiness index:} "A trust[worthiness] index is a composite and relative value that combines multiple trust indicators." \cite{ITUT2017Overview}
    \item \textit{Threat:} "Threat against a system refers to anything that can or may bring harmful effects to the state of the system and lead to improper service states." \cite{cho2019stram}
\end{itemize}



Self-localization is an essential service in IoT systems, creating dependencies to other components, services or entities \cite{li2020toward}.
Since UWB is used as the main technology for indoor localization \cite{coppens2022overviewUWB}, it is a perfect candidate for trustworthiness assessment. Our approach is threat-driven. Firstly, threats to the system are identified and mapped to the attributes of trustworthiness. This correspondence may be used to ensure that no aspect of trustworthiness is missed in the evaluation. From these threats, measurable quantities are identified that indicate the presence of a threat. Using this approach, \textit{meaningful} metrics are obtained (i.e., they correspond to realistic threats). Then, metrics are mapped to trustworthiness indicators in the value range from 0 to 1, with values below 0.5 being considered not trustworthy. The indicators are then combined in trustworthiness indices that represent the attributes of trustworthiness. By following this process, the contributions of this work are:
\begin{enumerate}
    \item A general framework for trustworthiness assessment that can be adapted to various IoT applications \bl{using the presented assessment as blueprint}. 
    \item A threat-driven metric selection and indicator computation to identify meaningful system measures.
    \item Experimental evaluation is conducted to provide insights into the strengths and weaknesses of the UWB self-localization service concerning trustworthiness, demonstrating that trustworthiness can improve the overall system performance.
\end{enumerate}
Note that the proposed approach focuses on the novelty of the methodology and does not claim completeness of a trustworthiness assessment.



\subsection{Defining Trustworthiness} \label{Sec: Defining Trustworthiness}

In this subsection, further key terms used throughout the article are introduced based on the previously given definitions. Since even the basic definitions vary in the literature, the focus is on capturing the essential characteristics of each definition.

\begin{figure}
    \centering
    \includegraphics[scale=1]{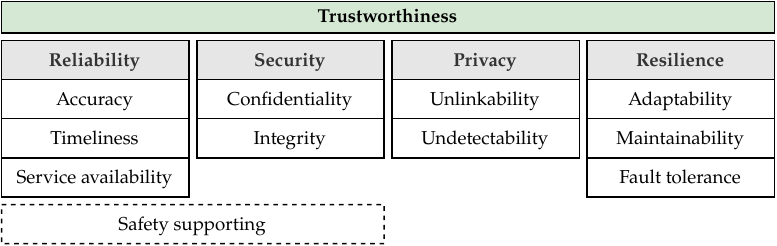}

    \caption{Taxonomy -- attributes and sub-attributes of trustworthiness.}
    \label{fig:taxonomy}
\end{figure}

Trustworthiness is divided into five main attributes (sometimes called pillars or characteristics) \cite{NIST2016cpsframework, buchheit30industrial}: safety, security, reliability, privacy and resilience. In \cref{fig:taxonomy}, a graphical summary of these attributes is given. The main focus of \textbf{safety} is to mitigate damage and harm to humans, objects, and the environment in which the system operates. Within the scope of this work (i.e., localization), safety is not considered a standalone attribute but is rather supported by reliability and security.

Traditionally, the primary goals of \textbf{security} are to protect \textit{confidentiality} (i.e., prevent unauthorized access), \textit{integrity} (i.e., prevent unauthorized alterations) and \textit{availability} (i.e., provide uninterrupted access to authorized subjects) of a system. For the purpose of this work, (service) availability is considered to fit better under the attribute of reliability, as the primary emphasis in security is on preventing malicious activities rather than merely ensuring operational functionality. \textbf{Reliability} is the ability of a system to provide a service under normal conditions, whereas \textbf{resilience} is the ability to adapt and recover from a state when a system is disrupted (e.g., by an attack). The related sub-attributes of reliability have the following meaning: \textit{accuracy} is a measure of the deviation of a measurement from the true value, \textit{timeliness} refers to the ability to deliver results within the required time frame. The sub-attributes of resilience are: \textit{adaptability} (i.e., ability to adjust to new conditions), \textit{maintainability} (i.e., the ease with which a system can be maintained and restored) and \textit{fault tolerance} (i.e., the capability to continue operating properly in the event of the failure).

\textbf{Privacy} refers to the right of an individual to control access to and confidentiality of their personally identifiable information. In this setting, the focus is on protecting users' location information and ranging data from unauthorized access or disclosure. \textit{Unlinkability} is a property ensuring that different transactions cannot be associated with a specific user. \textit{Undetectability} is an ability ensuring that a user's presence cannot be identified.

Note that the attributes may overlap. For instance, reliability and resilience are sometimes considered as subsets of availability. However, the focus of these attributes is different. Availability generally focuses on maximizing uptime, reliability deals with the probability of failures and resilience emphasizes speed of recovery.

\subsection{Related Work}
The concept of trustworthiness has been studied across various communication domains and substantial growth in relevance has recently been seen within the IoT and wireless sensor networks community. To provide guidelines for developing trusted communication infrastructure and services, the Telecommunication Standardization Sector of the International Telecommunication Union (ITU-T) has published a recommendation that discusses the concepts, provision, and evaluation processes of trustworthiness~\cite{ITUT2017Overview}. Although this report introduces several trustworthiness attributes, it lacks clarity on how these attributes contribute to the overall assessment process. A more comprehensive set of attributes contributing to system trustworthiness is proposed in~\cite{cho2019stram}. While key metrics for trustworthiness evaluation are discussed at the system level, the report does not offer quantitative approaches for assessing systems in operation. Focusing on industrial IoT applications,~\cite{IIOT2019trustworthiness} identifies five main trustworthiness attributes, namely reliability, security, privacy, resilience, and safety. The authors propose a generic framework for trustworthiness assessment based on these characteristics. However, they do not provide practical methods for implementing the assessment in specific industrial IoT applications, considering their unique requirements and limitations.

The notion of what constitutes a trustworthy system and how to assess the trustworthiness status of a system and its services highly depends on the specific application and the services the system offers. In the realm of localization applications, a few studies in the literature address trustworthiness evaluation from various perspectives. Considering only the reliability aspect of trustworthiness,~\cite{xu2011trust} proposed an algorithm that integrates a trustworthiness index to evaluate the reliability of the information reported by nodes, thereby mitigating the impact of faulty nodes on localization accuracy. \cite{kim2019novel} presents a blockchain-based trustworthiness evaluation and management model for wireless sensor networks. The authors defined trustworthiness metrics, e.g., honesty and intimacy, which can be computed based on measurements from high network layers, e.g., the number of successful and unsuccessful interactions and the time of interaction. \bl{These metrics are used to evaluate the trustworthiness of anchors, which the nodes rely on for localization. While evaluating anchors' trustworthiness is crucial, it is equally important to assess the trustworthiness of all other entities in the network, including the nodes themselves, as well as the overall system trustworthiness, using meaningful metrics.} \cite{jain2021simulation} presents a simulation framework focused on assessing the resilience of indoor ultrasound localization systems. However, their approach relies on metrics derived from localization error, which necessitates ground truth of true location—an impractical requirement in real-world implementations. A trustworthiness evaluation scheme for UWB communications is introduced in~\cite{peterseil2022datatrust}. This scheme evaluates reliability and security using machine learning (ML) techniques. However, its reliance solely on one metric, i.e., the channel impulse response, limits its ability to provide a holistic assessment of system trustworthiness.

\bl{In conclusion, existing studies such as~\cite{ITUT2017Overview,cho2019stram,IIOT2019trustworthiness} lack comprehensive quantitative assessment methods and practical implementation guidelines. They often focus narrowly on single aspects, like reliability in~\cite{xu2011trust}, or assess resilience with impractical requirements, such as ground truth localization in~\cite{jain2021simulation}. Other studies, such as~\cite{kim2019novel} and \cite{peterseil2022datatrust}, perform their trustworthiness assessment using limited metrics; [10] focuses only on anchors and higher network layers metrics, which may not be readily available or cover all trustworthiness attributes, and~[12] assesses reliability and security based solely on the channel impulse response. Both do not provide a holistic evaluation. Building upon these limitations found in the literature, a structured, practical, and general trustworthiness assessment framework is proposed. This framework can evaluate the state of all involved entities in the network, including the anchors, nodes, and the overall system, from the aspects of reliability, security, resilience, privacy and safety. It can also be easily adapted to new use cases and applications. In addition, the proposed method utilizes a handful of metrics covering these aspects of trustworthiness and is driven by measurement data from lower network layers, thereby enhancing the generality and agility of the scheme.}

\begin{figure}[t!]
    \centering
    \includegraphics[scale=1]{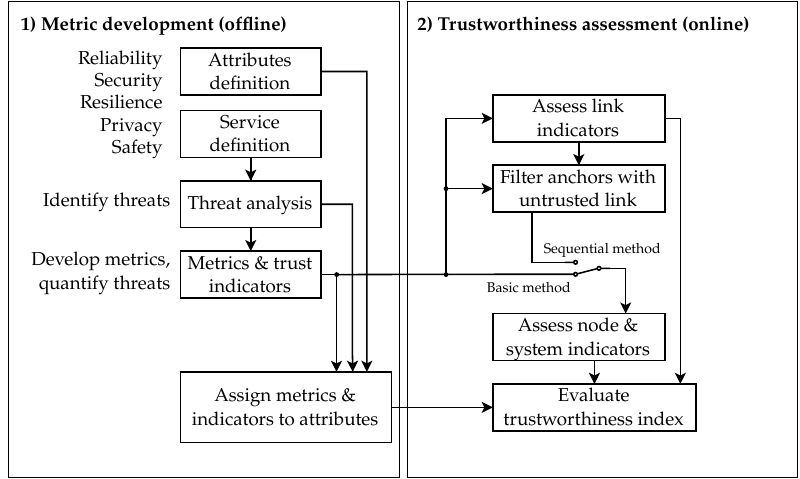}

    \caption{\bl{Methodology -- the metric development (left) is carried out once during the offline phase, while trustworthiness assessment (right) is used during the online phase.}}
    \label{fig:trust-process}
\end{figure}

\subsection{Methodology}

This study presents a methodology aiming to develop an application-centric trustworthiness assessment framework tailored specifically for UWB localization services. Figure~\ref{fig:trust-process} illustrates the proposed methodology workflow. \bl{By running through this workflow, trustworthiness can also be assessed for other IoT use cases, with the presented one serving as a blueprint.} The proposed methodology consists of two main phases, including metric development
(Figure~\ref{fig:trust-process}, left) and trustworthiness assessment (Figure~\ref{fig:trust-process}, right).
The final output is the trustworthiness index of the system.

\bl{
The metric development phase links the general concept of trustworthiness to measurable quantities in the specific application. It consists of the following steps:
\begin{enumerate}[label=(\roman*)]
    \item \textit{Trustworthiness attributes}: Trustworthiness is defined as a holistic measure that signalizes if the system is working as intended. Hence, it first requires an understanding of which aspects of the system have to be observed. As proposed in literature \cite{NIST2016cpsframework, buchheit30industrial}, the currently considered system attributes include reliability, security, privacy, resilience and safety (c.f. \cref{Sec: Defining Trustworthiness}).
    \item \textit{Service definition}: To match the trustworthiness attributes to UWB self-localization, a clear definition of the operational principle is required. This step establishes a clear and comprehensive understanding of the system under evaluation, laying the groundwork for the subsequent step of threat analysis.
    \item \textit{Threat analysis}: The threat analysis is a critical step aimed at identifying potential vulnerabilities of the defined service. It leverages critical parameters of the system at hand. The most challenging aspect is to address the entire set of identified trustworthiness attributes.
    \item \textit{Metrics \& trust indicators}: While there are many possible measurable service parameters (referred to as metrics), this step identifies those that are relevant to detecting the likelihood of a threat. To make relevant metrics comparable, they have to be further mapped to trustworthiness indicators with a value range in the interval [0,1] where 0 and 1, respectively, indicate not trustworthy and trustworthy.
    \item \textit{Assign metrics \& indicators to attributes}: In order to obtain a measure for each trustworthiness attribute, first, the identified threats have to be assigned to the corresponding trustworthiness attributes. This establishes a coherent mapping between the quantitative evaluation metrics and the qualitative trust attributes. Finally, the indicators are combined to higher level indices that provide a high-level trustworthiness assessment.
\end{enumerate}
}
\noindent
Depending on which entity in the localization service is measured, the trustworthiness indicators are categorized into \textbf{node} (referring to the state of the node), \textbf{link} (referring to the state of the link between the node and each anchor), and \textbf{system} (referring to the state of whole localization system) indicators.
\cref{sec:threat-analysis}, \cref{sec:metrics-trustworthiness-indication} and \cref{sec:selected-trustworthiness-indicators}, respectively, describe threat analysis, metric derivation and assignment in detail.

The second phase of the proposed method is trustworthiness assessment, where the system's trustworthiness is evaluated online. This evaluation involves monitoring defined indicators and combining them to obtain a unified trustworthiness index for each attribute, as well as an overall trustworthiness index for the entire system over time. Trustworthiness is assessed based on link, node, and system indicators. Two methods are proposed, called basic and sequential methods. In the basic method, trustworthiness evaluation is conducted based on all metrics and their corresponding indicators which are monitored in the system. In contrast, the sequential method involves filtering out anchors with untrusted links and then updating the node, link and system indicators. This approach would eventually enhance the robustness of the localization service, as will be shown in the results. The results will reveal that considering trustworthiness not only provides valuable insights into the system's status but can also enhance its overall performance. \bl{In terms of scalability and deployment, trustworthiness is assessed locally on a node in decentralized manner. The complexity does, therefore, not increase with the number of nodes present in the network and linearly with the number of available anchors.}

\section{Service Definition}\label{sec:model}
A 2D self-localization service of one battery powered node in an environment with multiple cable-powered anchors is considered. The service relies on UWB range measurements and subsequently processes the distance estimates (ranges) to a location estimate. Finally, the location estimate can serve an application as a functional basis.


\subsection{Range-Based Self-Localization}\label{sec:operational_principle}

\begin{figure}
    \centering
    \begin{tabular}{cc}
       \input{texfig/DSTWR_CIR} &   \input{texfig/packet_structure} \\[3mm]
         (a) & \hspace{0.5cm} (b)
    \end{tabular}

    \caption{(a) The double-sided two-way ranging message exchange using four packets with channel measurements; the superscripts are omitted for simplicity, e.g., $t_\mathrm{a}$ is used instead of $t_\mathrm{a}^{(A)}$. (b) UWB packet configuration possibilities and the position of timestamping within the packet according to the IEEE 802.15.4 specification.}
    \label{fig:ranging_and_packet}
\end{figure}
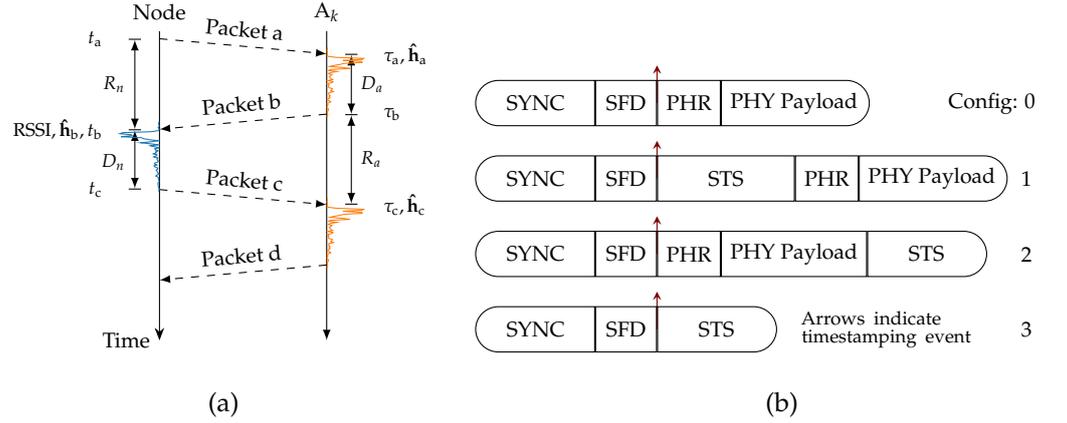

The node performs ranging with a subset $\mathcal{A}_\text{eval} = \{\text{A}_{1}, \ldots, \text{A}_{K}\} \subseteq \mathcal{A}$ of all existing anchors $\mathcal{A}$. The subset is selected either through the communication range of the node or through another criterion introduced later in \ref{sec:assessment-anchor-selection}.
For ranging, three packets have to be exchanged and the measured time intervals are converted into a range estimate. This packet exchange is referred to as \textit{double-sided two-way ranging} \cite{neirynck2016alternative}.  \bl{In \cref{fig:ranging_and_packet}(a), the packet exchange cycle between node and anchor $A \in \mathcal{A}$ with the measured timestamps at the node $t_\mathrm{a}^{(A)}, t_\mathrm{b}^{(A)}, t_\mathrm{c}^{(A)}$ and the anchor $\tau_\mathrm{a}^{(A)}, \tau_\mathrm{b}^{(A)}, \tau_\mathrm{c}^{(A)}$, and channel impulse responses $\hat{\mathbf{h}}_\mathrm{a}^{(A)}, \hat{\mathbf{h}}_\mathrm{b}^{(A)}, \hat{\mathbf{h}}_\mathrm{c}^{(A)}$ is illustrated. Based on the round trip intervals, $R_a^{(A)}=\tau_\mathrm{c}^{(A)}-\tau_\mathrm{b}^{(A)}$ and $R_n^{(A)}=t_\mathrm{b}^{(A)}-t_\mathrm{a}^{(A)}$, and the response delays, $D_a^{(A)}=\tau_\mathrm{b}^{(A)}-\tau_\mathrm{a}^{(A)}$ and $D_n^{(A)}=t_\mathrm{c}^{(A)}-t_\mathrm{b}^{(A)}$, the ranges are computed} according to \cite{neirynck2016alternative} by

\begin{equation}
    \hat{r}^{(A)} = c\,\frac{R_a^{(A)} R_n^{(A)}-D_a^{(A)} D_n^{(A)}}{R_a^{(A)} +D_a^{(A)} +R_n^{(A)} +D_n^{(A)}} \, , \label{eq:DSTWR}
\end{equation}
where $c$ is the speed of light. \bl{Processing time-of-flight according to \eqref{eq:DSTWR} is known as \textit{asymmetric} double-sided two-way ranging. The method provides implicit synchronization of node and anchor and does not impose constraints on the response delays $D_n^{(A)}$ and $D_a^{(A)}$. Detailed derivation can be found in the reference.}
%
The anchor $A \in \mathcal{A}$ is located at known position $\mathbf{x}^{(A)}\in \mathbb{R}^2$. The ranges are collected in $\mathbf{\hat{r}}$, and the anchor positions in $\mathbf{X}$, respectively,

\begin{align*}
    \mathbf{\hat{r}} &=
    \left[
        \hat{r}^{(\mathrm{A}_1)}, \ldots, \hat{r}^{(\mathrm{A}_K)}
    \right]^\top \in \mathbb{R}^K\, , 
    ~~~~~\mathbf{X} =
    \left[
        \mathbf{x}^{(\mathrm{A}_1)} ~\text{\scriptsize\rotatebox[origin=c]{90}{$---$}}~ \dots ~\text{\scriptsize\rotatebox[origin=c]{90}{$---$}}~ \mathbf{x}^{(\mathrm{A}_K)}
    \right] \in \mathbb{R}^{2 \times K} \,. \label{eq:Xvec}
\end{align*}

Anchor $A$ uses the payload of packet b to share $\tau_\mathrm{a}$, $\hat{\mathbf{h}}_\mathrm{a}$ and $\tau_\mathrm{b}$ with the node. To provide it with $\tau_\mathrm{c}$ and $\hat{\mathbf{h}}_\mathrm{c}$, a fourth packet is used.

For each localization sequence, the node first carries out ranging with all available anchors. Then, the node estimates its position $\mathbf{x}$ by computing
\begin{equation}
    \hat{\mathbf{x}} = f_\text{loc}\left( \mathbf{\hat{r}},\mathbf{X}\right)\,,
    \label{eq:localization}
\end{equation}
where the localization function $f_\text{loc}$ can be chosen according to the system requirements. Here, a simple least-squares localization \cite{cheng2012localizationsurvey} is considered.
The channel impulse responses and other channel-related features (e.g., the received signal strength indicator (RSSI)), are simultansously recorded at the transceiver and frequently used for non line-of-sight detection \cite{DW1000APS06}.

\subsection{UWB Packet Structure}
\bl{According to the 802.15.4-2011 standard \cite{coppens2022overviewUWB}, the exchanged packets consist of four fields: the SYNC and SFD which together form the synchronization header (SHR), the physical layer header (PHR) and PHY payload field for data transmission (see \cref{fig:ranging_and_packet} (b), config 0).
Within the packet, a timestamping event is defined at the end of the SHR. It serves as a reference point for measuring the time intervals needed for \eqref{eq:DSTWR}.
The exchanged physical layer (PHY) packets have one of the logical structures shown in \cref{fig:ranging_and_packet}(b), selected through node configuration. Cconfigurations 0, 1, and 2 enable payload data to be appended, while configuration 3 serves purely for time measurements. Moreover, configurations 1, 2 and 3 include the scrambled timestamp sequence (STS) field, which adds an additional security mechanism. The use of the STS requires common knowledge of the keys and cryptographic scheme between transmitter and receiver. The location of the STS may vary depending on the configuration (see \cref{fig:ranging_and_packet}(b), config 1-3).}

\subsection{Receive Time Estimation for Ranging}

Time measurements used for ranging refer to the moment the timestamping event occurs, i.e., when the end of the SHR appears at the antenna.
The timestamp at packet reception is based on a leading-edge detection algorithm denoted by $f_\text{LDE}$, i.e.,
\begin{equation*}
    \text{RX\_STAMP} = f_\text{LDE}\big(\text{channel},\text{SHR})\,. \label{eq:RX_STAMP}
\end{equation*}
While currently no UWB transceiver manufacturer discloses information about its leading-edge detection implementation, it is understood that this algorithms rely either on the received SHR or on the STS waveform, and that its accuracy is influenced by the propagation channel.

Implementations using the SHR sequences for timestamping are vulnerable to distance spoofing attacks. This can be prevented by using the STS sequence for timestamping, which is derived from a cryptographic function depending on a key only known to legitimate devices. However, \cite{leu2022ghost} demonstrated that injection of a random signal with high power at the time the legitimate transmitter sends the sequence could still cause distance reduction in some cases since the sequence seems not to be evaluated bit-wise, but only by correlation of the received signal with a template of the sequence.
\subsection{Protocol Stack for Data Exchange}
Communication is encoded in medium access control (MAC) frames as defined by IEEE802.15.4. The MAC layer offers two addressing modes: extended unique IDs and short IDs that are dynamically assigned upon association with a private area network. The standard also specifies an authenticated encryption with associated data method for MAC layer security, providing payload data confidentiality, authenticity, and replay protection. Additionally, a MAC frame can request an acknowledgment from the receiver to confirm the correct packet reception, and if not received, this will trigger a retransmission.

\subsection{Known Exploits for UWB Ranging}
As UWB ranging is used in security-relevant applications, it is a target for attacks. Examples include distance reduction attacks for keyless entry systems. Attacks occur on the node level (e.g., targeted battery drain), on the link level (e.g., early detect/late commit attacks), or on the system level (e.g., anchor node impersonation). Therefore, link level attacks are specifically studied \cite{peterseil2023,poturalski2010cicada,singh2021security}.


\section{Threat Analysis}\label{sec:threat-analysis}
In this section, some major threats to the UWB systems described in \cref{sec:model} are classified.
Primarily, the negative events that can disrupt one or more basic trustworthiness attributes of such systems are discussed. The threats can be divided into two main categories: \textit{native threats} to self-localization (i.e., any system that uses UWB localization is subject to these threats) and \textit{application-specific} ones (i.e., these threats depend on the broader context of a given application). The main focus of the paper is laid on the native threats.

Furthermore, the threats specific to nodes (i.e.,\ the node itself), links (i.e., communication between a node and an anchor) and a system (i.e., localization service) are distinguished. The detailed overview is provided in \cref{tab:node_threats} for node threats, in \cref{tab:link_threats} for link threats and in \cref{tab:system_threats} for system threats. The node threats (e.g.,\ overheating) may apply to anchors as well. Since the focus is on self-localization, the threats affecting anchors are manifested as link threats (e.g.,\ weak signal) or as system threats (e.g., not enough threats). Note that the lists in the tables are not meant to be complete and the threat analysis depends on a specific use case. Furthermore, some threats can overlap or propagate. For example, software failure of an anchor could cause a weak signal (link), which in turn might result in a shortage of anchors (system). Moreover, the application-specific threats can be implied by the native threats (e.g., in a robot-assisted warehouse, one malfunctioning anchor could affect the accuracy of the localization service, which might lead to collisions of robots with humans).

\begin{table}[H]
\caption{An overview of node threats.\label{tab:node_threats}}
\rowcolors{2}{gray!10}{white}
\begin{tabular}{>{\raggedright\arraybackslash}p{0.2\textwidth}
                >{\raggedright\arraybackslash}p{0.35\textwidth}
                >{\raggedright\arraybackslash}p{0.35\textwidth}}
    \toprule
    \textbf{Threats} & \textbf{Examples} & \textbf{Impact} \\
    \midrule
    \textit{Hardware and software failures} &
    Software crashes, invalid configuration, firmware corruption, physical destruction, harsh environmental conditions &
    Node downtime, node malfunction, additional maintenance costs \\

    \textit{Overheating} &
    Misconfigurations, bugs, poor ventilation &
    Decreased ranging accuracy, higher power consumption, fire hazard, shorter lifespan of the device \\

    \textit{Low battery} &
    Incorrect power consumption settings, environmental conditions, battery aging, mismanagement of recharging&
    Decreased anchor performance, anchor downtime \\
    \bottomrule
\end{tabular}
\end{table}

\begin{table}[H]
\caption{An overview of link threats.\label{tab:link_threats}}
\rowcolors{2}{gray!10}{white}
\begin{tabular}{>{\raggedright\arraybackslash}p{0.2\textwidth}
                >{\raggedright\arraybackslash}p{0.35\textwidth}
                >{\raggedright\arraybackslash}p{0.35\textwidth}}
    \toprule
    \textbf{Threats} & \textbf{Examples} & \textbf{Impact} \\
    \midrule
    \textit{Channel obstructions} &
    Reflections, non line-of-sight &
    Decreased ranging accuracy \\

    \textit{Weak signal} &
    Large distance between an anchor and a node &
    Higher packet error rate, data throughput \\

    \textit{Interference} &
    Unintentional interference (in-band or out-band), jamming &
    Decreased ranging accuracy, denial of service, increased error rates \\

    \textit{Active attackers} &
    Jamming, packet injection, preamble tampering, active probing, denial of service, payload overwriting &
    Compromised security and reliability \\
    \bottomrule
\end{tabular}
\end{table}

\begin{table}[H]
\caption{An overview of system threats.\label{tab:system_threats}}
\rowcolors{2}{gray!10}{white}
\begin{tabular}{>{\raggedright\arraybackslash}p{0.2\textwidth}
                >{\raggedright\arraybackslash}p{0.35\textwidth}
                >{\raggedright\arraybackslash}p{0.35\textwidth}}
    \toprule
    \textbf{Threats} & \textbf{Examples} & \textbf{Impact} \\
    \midrule
    \textit{Improper anchor configuration} &
    Incorrect anchor placement &
    Decreased localization accuracy, degraded system performance \\

    \textit{Not enough anchors} &
    Insufficient number of anchors for unambiguous localization &
    Localization service fails \\

    \textit{Eavesdropping} &
    A passive attacker with a UWB receiver &
    Compromised confidentiality, privacy breaches of localization data \\

    \textit{Evil anchors} &
    Impersonating anchors and announcing wrong time information or wrong anchor position &
    Compromised security and reliability \\

    \bottomrule
\end{tabular}
\end{table}

\section{Metrics and Trustworthiness Indication}\label{sec:metrics-trustworthiness-indication}
\begin{figure}[t!]
    \centering
    \includegraphics[width=0.9\textwidth]{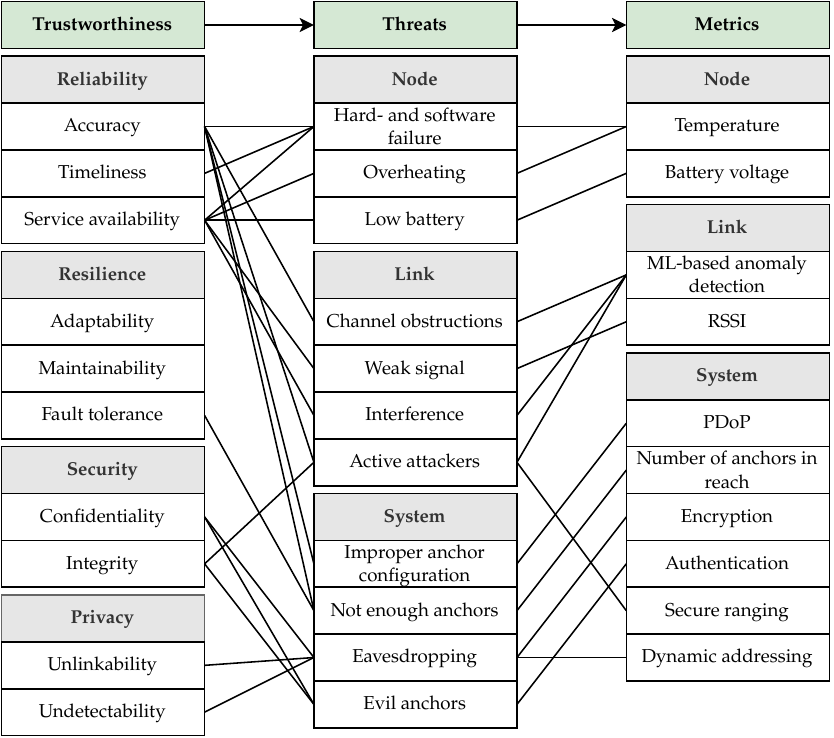}

    \caption{Mapping of trustworthiness attributes to metrics through threats.}
    \label{fig:mapping}
\end{figure}

Metrics are used to evaluate the characteristics of a system in the respective attributes. While the attribute definition alone is too general to determine meaningful and relevant metrics, metrics that enable the detection of the previously introduced threats are chosen. The approach of mapping metrics to attributes for the discussed UWB self-localization system is depicted in \cref{fig:mapping}.

As each chosen metric has a different dynamic range, they are a priori not comparable. Thus, a mapping to trustworthiness indicators with specific conditions is required. To further extract high-level evaluation (e.g., the trustworthiness with respect to reliability), such indicators can be unified into trustworthiness indices.

In this section, the general notation of metrics, their mapping to trustworthiness indicators and their unification to trustworthiness indices is provided. In the following section, metrics are discussed one by one.

\subsection{Metrics}
In this work, measurable quantities that are available from the UWB transceiver or from intermediate information within the localization scheme are selected. Those metrics, which are detailed in the subsequent section, are: temperature (temp), battery voltage (bat), ML-based anomaly detection \cite{peterseil2022datatrust} (ml), RSSI (rssi), position dilution of precision (PDoP, pdop), number of anchors in reach (na), encryption used (enc), authentication used (auth), secure ranging used (sr), dynamic addressing used (da). Metrics are denoted by $m_i$ where $i$ is the abbreviation of the metric (e.g., $m_\text{rssi}$ for RSSI) or just by $m$.

For later use, the following sets according to measurements that relate to the node (temp and bat), to the link with an anchor (ml and rssi) or to the system configuration (all remaining metrics) are defined:
\begin{align*}
    \mathcal{M}_\textit{node} & = \left\{ m_{\mathrm{temp}}, m_\mathrm{bat} \right\} & &\text{node metrics} \\
    \mathcal{M}_\textit{link} & = \left\{ m_\mathrm{ml},
m_\mathrm{rssi} \right\} & &\text{link metrics} \\
    \mathcal{M}_\textit{sys} & = \left\{ m_\mathrm{pdop},
m_\mathrm{na}, m_\mathrm{enc}, m_\mathrm{auth}, m_\mathrm{sr}, m_\mathrm{da} \right\} & & \text{system metrics} \\
\mathcal{M} & = \mathcal{M}_\textit{node} \cup \mathcal{M}_\textit{link} \cup \mathcal{M}_\textit{sys} & &\text{all metrics}
\end{align*}
For $m\in\mathcal{M}_\text{link}$, $m^{(A)}$ denotes the metric measurement between the node and the anchor $A\in\mathcal{A}_\text{eval}$. Note that a metric can be either real valued  $m\in \mathbb{R}$ (e.g., temperature readings), or binary with $m\in \{\text{state 1}, \text{state 2}\}$ (e.g., encryption on/off).

\subsection{Mapping to Trustworthiness Indicators}

\begin{figure}
    \centering

    \begin{tabular}{@{}c@{}c@{}}
          \includegraphics[scale=1]{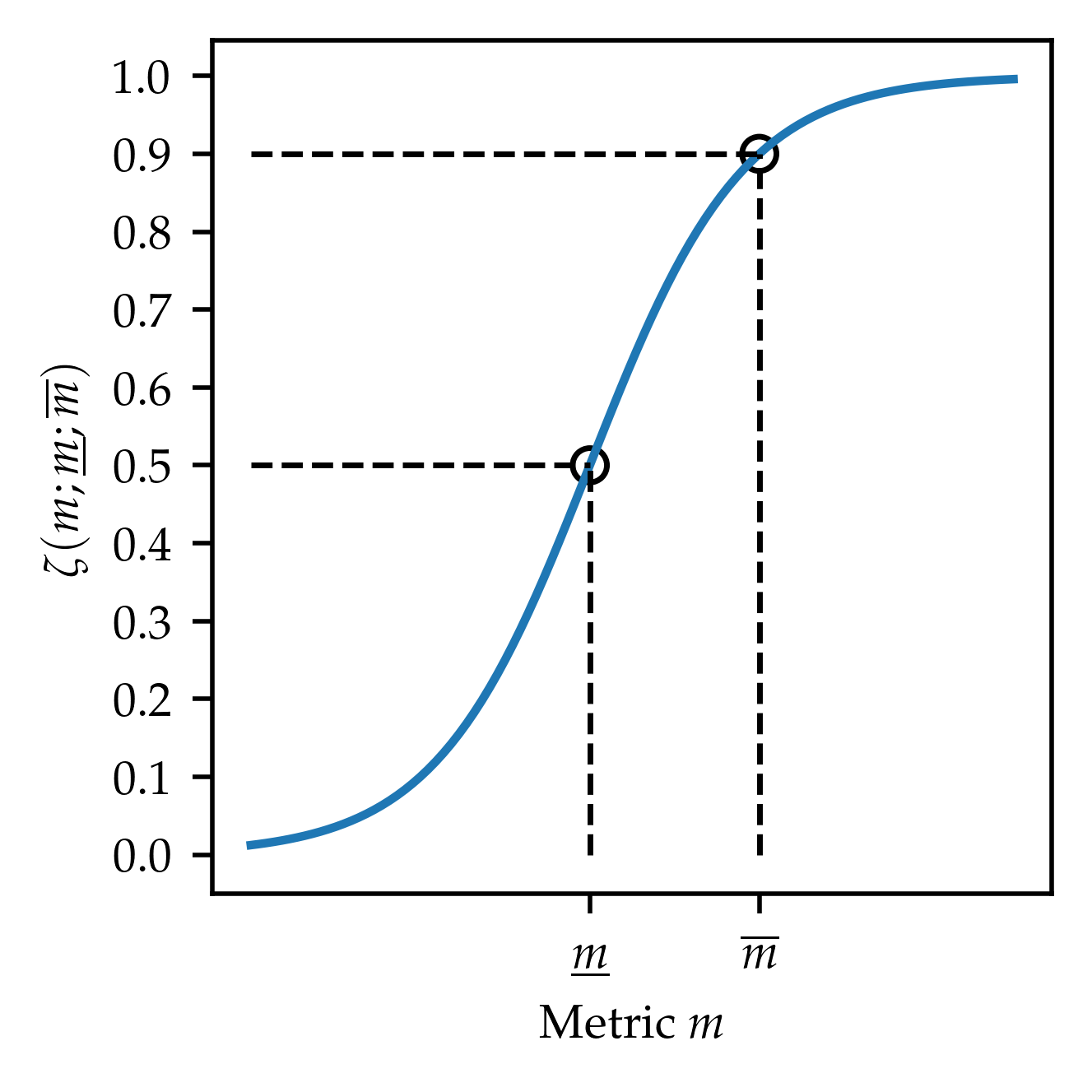}
          &
          \includegraphics[scale=1]{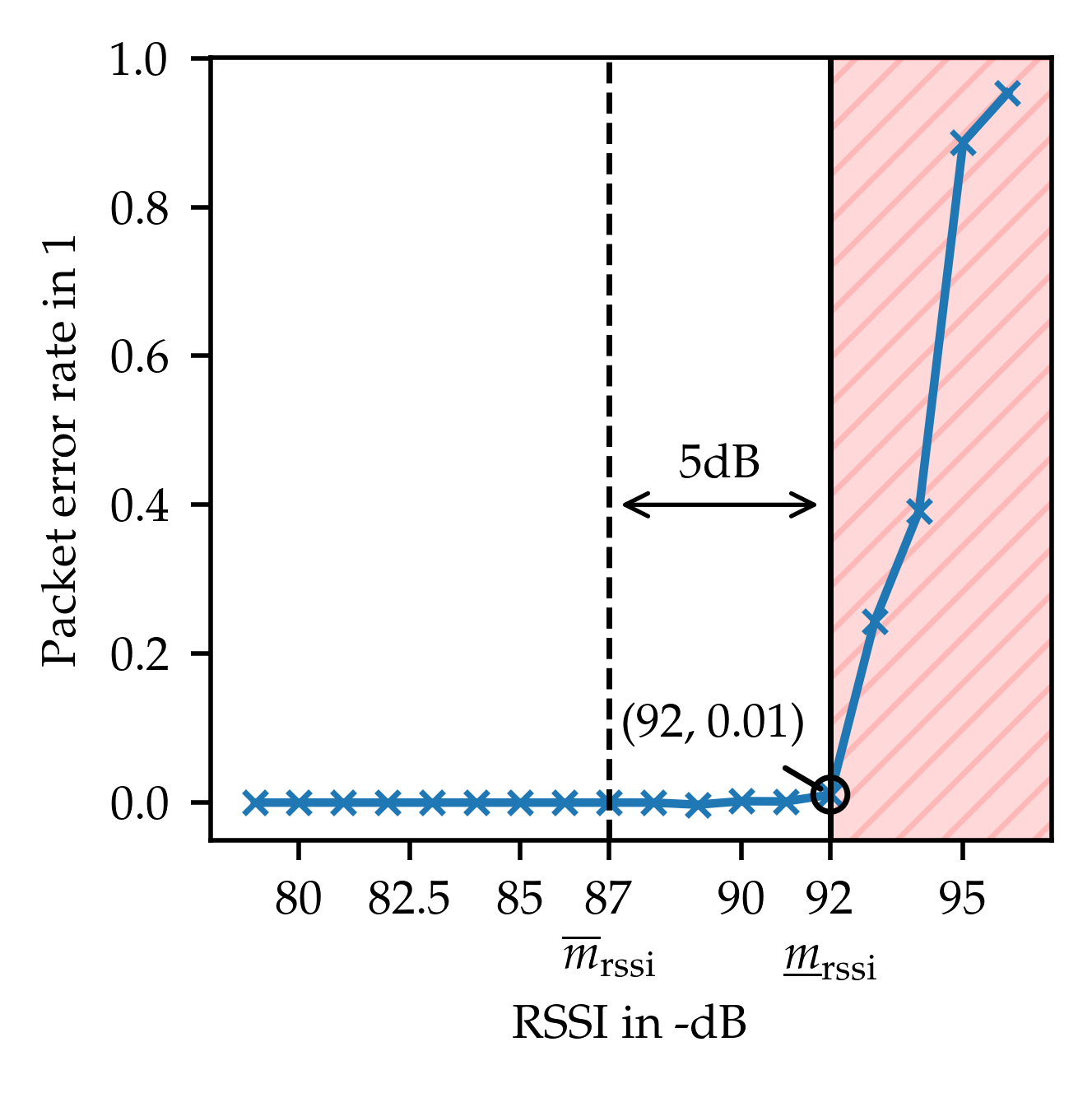}
          \\
         (a) & 
         (b)
    \end{tabular}

    \caption{(a) The sigmoid function is used to map metrics to a trustworthiness indicator. (b) The RSSI is used to account for anchors with low signal strength that might not respond.}
    \label{fig:sigmoid-rssi}
\end{figure}

In this step, the selected metrics $m \in \mathcal{M}$ are mapped to unified trustworthiness indicators $T_m$ bound to the interval [0,1].
To simplify the notation of superscripts and subscripts, e.g., $T_{\text{ml}}$ is written instead of $T_{m_\text{ml}}$. Similarly, for $A \in \mathcal{A}$,  $T_{\text{ml}^{(A)}}$ is used instead of $T_{m_\text{ml}^{(A)}}$. This normalization is required to account for the individual sensitivity of the metrics and their application-specific importance. For \textbf{real value metrics}, a sigmoid function
\begin{equation}
    T_m = \zeta\left(m; \underline{m}, \overline{m}\right) =
        \frac{1}{1+\mathrm{e}^{-g\frac{1}{\overline{m}-\underline{m}}\left(m-\underline{m}\right)}}\, , \label{eq:trustworthiness_indicator_mapping}
\end{equation}
with the constant $g=\ln{9}$ is used for the mapping, c.f. \cref{fig:sigmoid-rssi}(a). The sensitivity can be adjusted by tuning parameters, such that $\underline{m}$ marks the transition from not trustworthy to trustworthy at $T_{{m}}=0.5$, while $\overline{m}$ is set to a value representing a reasonable level of trust, i.e., $T_{{m}}=0.9$.
\bl{The sigmoid function enables the appropriate mapping $\zeta : \mathbb{R} \rightarrow [0, 1] $, ensuring that any real-valued input is transformed into a value within the interval [0,1]. Furthermore, the sigmoid function provides a non-zero gradient even in saturated regions, i.e., where $T_{{m}}>0.9$ or $T_{{m}}<0.1$. This characteristic is beneficial when the proposed framework is used in the context of trustworthiness management to account for changes in trustworthiness over time.}
\textbf{Binary metrics} are either mapped to 1 if they are considered trustworthy or to 0, otherwise.


\subsection{Trustworthiness Index}
To obtain a unified trustworthiness index for each attribute (i.e., reliability, resilience, security, and privacy), the trustworthiness indicators must be combined. Here,  trustworthiness is defined by the least trusted component. Therefore, the minimum function is chosen as the evaluation criterion. For $m\in\mathcal{M}_\text{link}$
\begin{equation}
    T_{m}^{*} = \min\{ T_{m^{(A)}} ~|~ A\in\mathcal{A} \}\,. \label{eq:TWindex_link}
\end{equation}
The trustworthiness index per attribute is given by
\begin{align*}
    I_\text{rel} & = \min \{T_\text{temp},T_\text{ml}^{*},T_\text{pdop},T_\text{sec},T_\text{bat},T_\text{rssi}^{*},T_\text{na}\} \\
    I_\text{res} & = \min \{ T_\text{na} \} \\
    I_\text{sec} & = \min \{ T_\text{enc},T_\text{da},T_\text{aut},T_\text{sec},T_\text{ml}^{*} \}  \\
    I_\text{priv} & = \min \{ T_\text{da}\}  \\
\end{align*}
Note that the corresponding trustworthiness indicators are found through the relations in
\cref{fig:mapping}. Reliability consists of Accuracy, Timeliness and Service availability.
Accuracy links to the threats Hard- and software failure, channel obstruction, active attackers, improper anchor configuration and not enough anchors. The corresponding metrics (and hence indicators) are temperature, ML-based anomaly detection, PDoP and number of anchors. Similarly, Indicators for timeliness and service availability are found.
Most notably, the same metrics can be used for quantification of different attributes.

An overall trustworthiness index 
is given by
$$I = \min \{I_\text{rel}, I_\text{res}, I_\text{sec}, I_\text{priv}\}.$$

\subsection{Trustworthiness Enhanced Anchor Selection Scheme}\label{sec:assessment-anchor-selection}
Conventionally, self-localization is performed with all anchors in communication range to the node (here referred to as \textit{basic} approach). However, intermediate results of link level trustworthiness indicators \eqref{eq:TWindex_link} may be directly used to select only anchors that possess trustworthy links with the node (here referred to as \textit{sequential} approach).
This distinction is formalized by the definition of $\mathcal{A}_\mathrm{eval}$ (c.f. \cref{sec:operational_principle}) with
\begin{align*}
    \mathcal{A}_\mathrm{eval} =
\begin{cases}
\mathcal{A} & \text{if the basic method is used,} \\
\Big\{ A \in \mathcal{A} ~|~ \min\limits_{m \in \mathcal{M}_\mathrm{link}} T_{m^{(A)}} \geq 0.5\Big\} & \text{if the sequential method is used.}
\end{cases}
\end{align*}





\section{Selected Trustworthiness Indicators}\label{sec:selected-trustworthiness-indicators}

The metrics that are linked to the threats, as depicted in \cref{fig:mapping}, are now introduced.
Furthermore, by inspecting the metric values, the thresholds required for mapping to the trustworthiness indicator in \eqref{eq:trustworthiness_indicator_mapping} are derived.

\subsection{Temperature (\texorpdfstring{$T_\mathrm{temp}$}{Ttemp})}
Temperature is used to detect overheating from external sources or to indicate hardware- or software failures.
The maximum operation temperature specified in the UWB-transceiver datasheet is used as mapping parameter $\underline{m}_\mathrm{temp} = \SI{85}{\celsius}$. The second parameter is set to $\overline{m}_\mathrm{temp} = \underline{m}_\mathrm{temp} - \SI{10}{\celsius}= \SI{75}{\celsius}$.

\subsection{Battery Voltage (\texorpdfstring{$T_\mathrm{bat}$}{Tbat})}
\begin{figure}
    \centering

    \begin{tabular}{@{}c@{}c@{}}
          \includegraphics[scale=1]{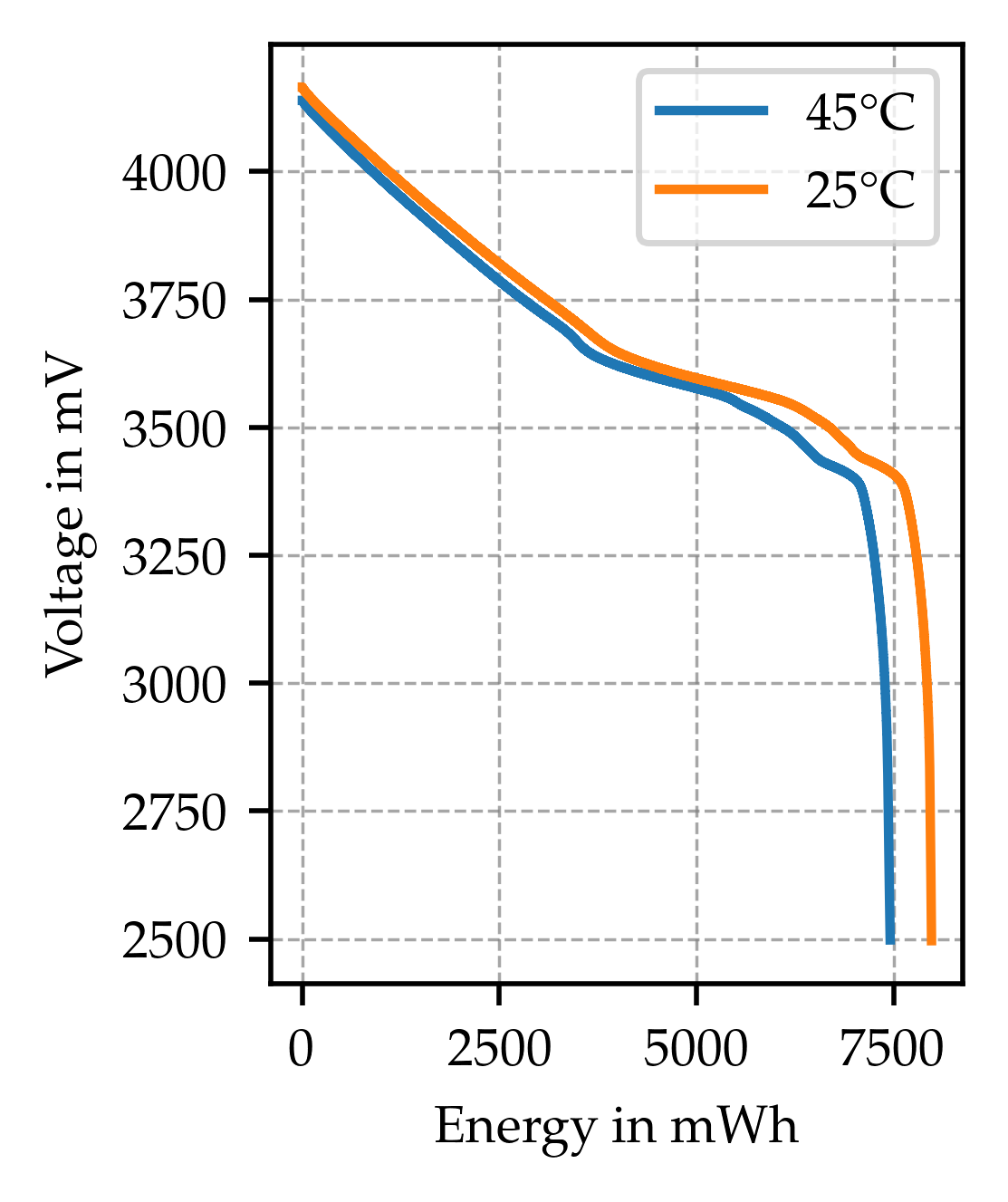}
          &
          \includegraphics[scale=1]{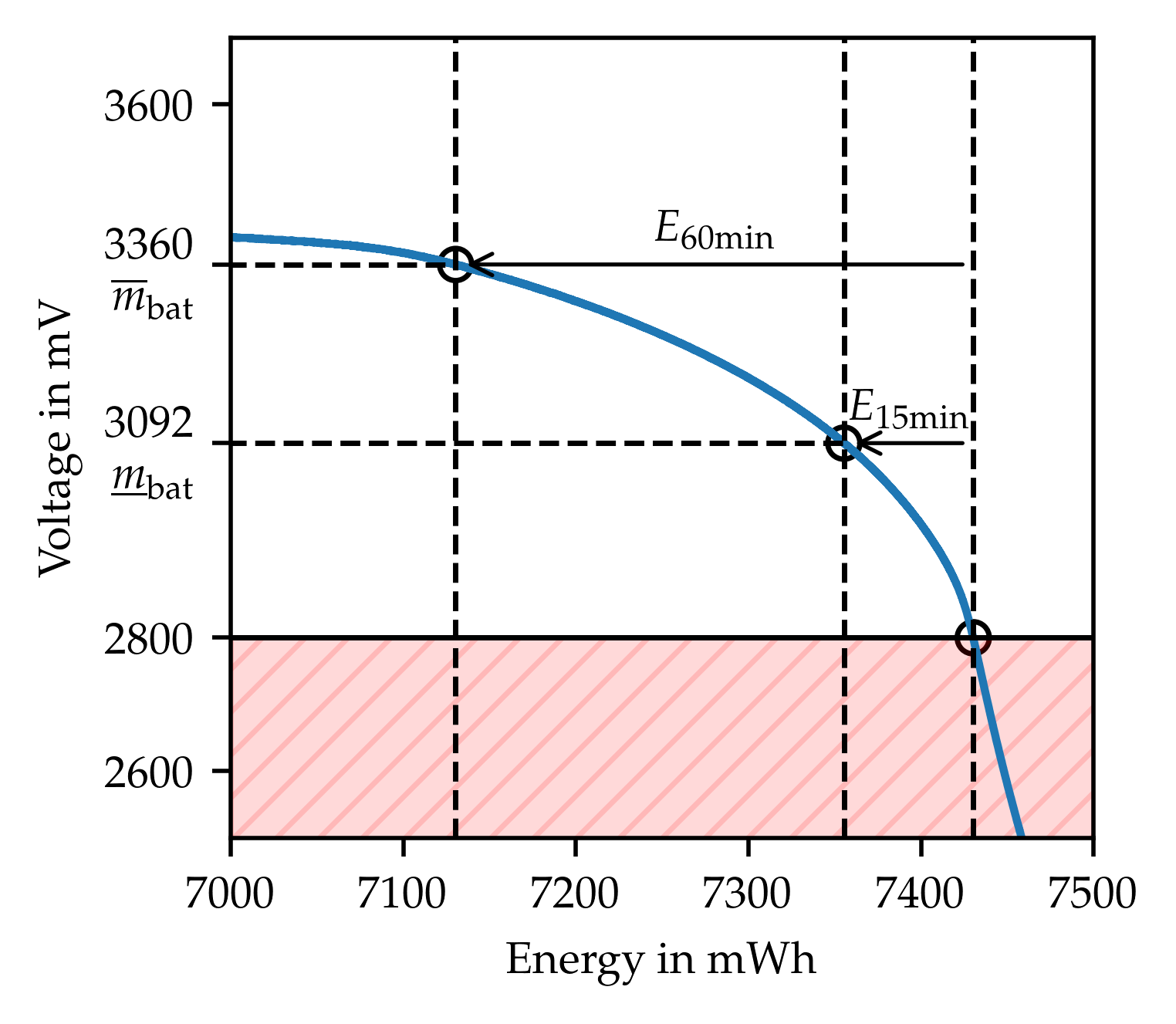}
          \\
         (a) & 
         (b)
    \end{tabular}

    \caption{(a) Discharge curve of an INR 18650-20R type battery \cite{ZHENG2016513} used to supply the node. (b) The discharge curve for \SI{45}{\celsius} is used to derive tuning parameters for mapping the battery voltage to the corresponding trustworthiness indicator.}
    \label{fig:discharge}
\end{figure}

Battery voltage is used as a metric to indicate the low battery threat and, hence, issues with the service availability.
Through discharge curves of batteries captured for the typical node current, c.f. \cref{fig:discharge}(a), the remaining battery time can be estimated \cite{XING2014106}. Battery voltage measurements are required as input to conclude the current battery charge.

The minimum operation voltage of the node, \SI{2.8}{\volt}, can be used to find the tuning parameters $\underline{m}_\mathrm{bat}$ and $\overline{m}_\mathrm{bat}$. They were set according to allow further operation for time spans of \SI{15}{\minute} and \SI{60}{\minute}, respectively. 
This is achieved by first estimating the corresponding amounts of energy consumed by the node $E_\mathrm{15min}$ and $E_\mathrm{60min}$. In \cref{fig:discharge}(b), from the intersection of the minimum operation voltage with the discharge curve, the curve is traced back by $E_\mathrm{15min}$ and $E_\mathrm{60min}$ to get the corresponding voltage levels used as tuning parameters. They are found to be  $\underline{m}_\mathrm{bat} = \SI{3092}{\milli\volt}$ and $\overline{m}_\mathrm{bat} = \SI{3360}{\milli\volt}$. The discharge curve for \SI{45}{\celsius} is used to account for the increased temperature with respect to ambient temperature during operation.

\subsection{ML-based Anomaly Detection (\texorpdfstring{$T_\mathrm{ml}$)}{Tml}}
\begin{figure}
    \centering

    \begin{tabular}{@{}c@{}c@{}}
          \includegraphics[scale=1]{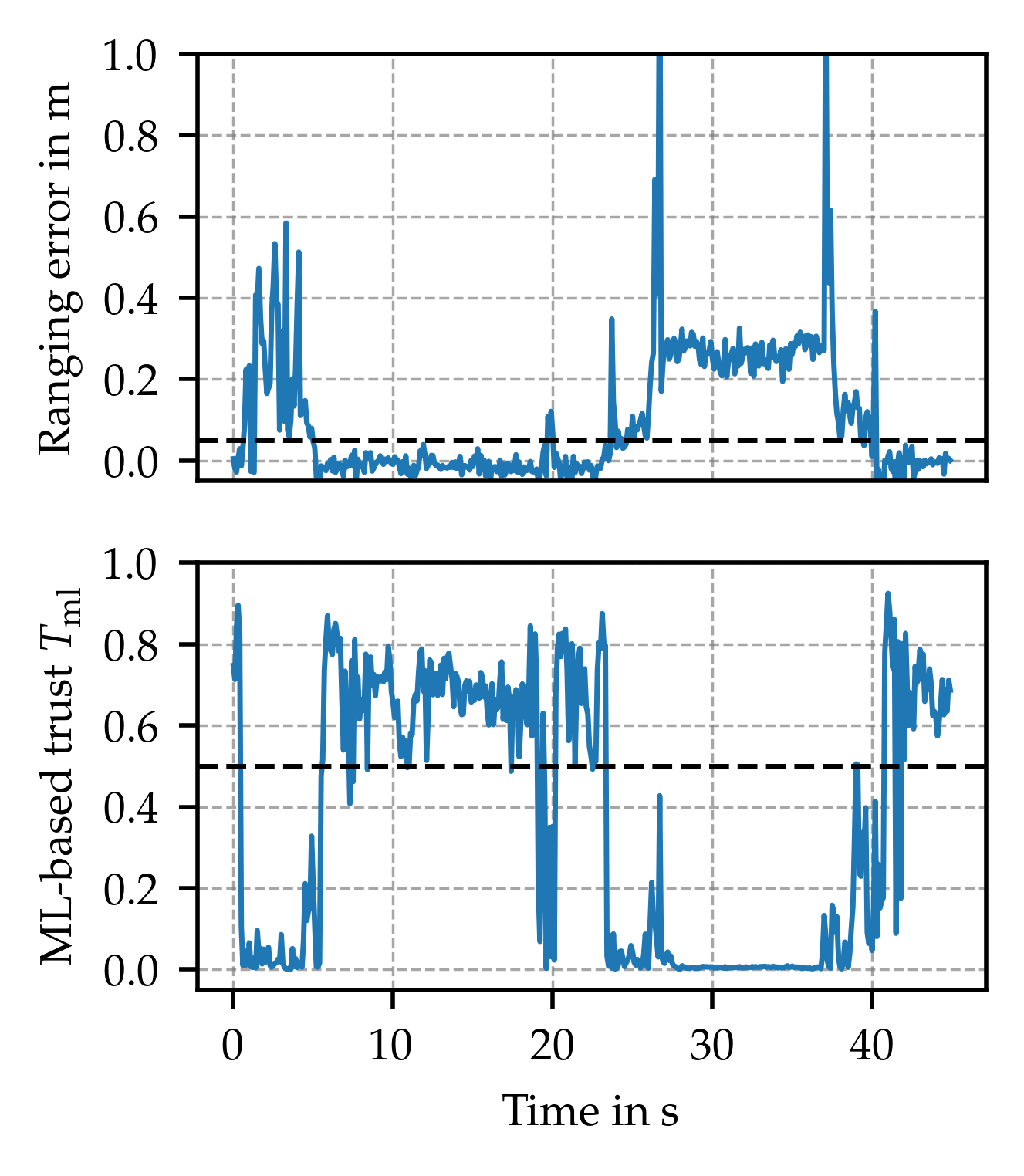}
          &
          \includegraphics[scale=1]{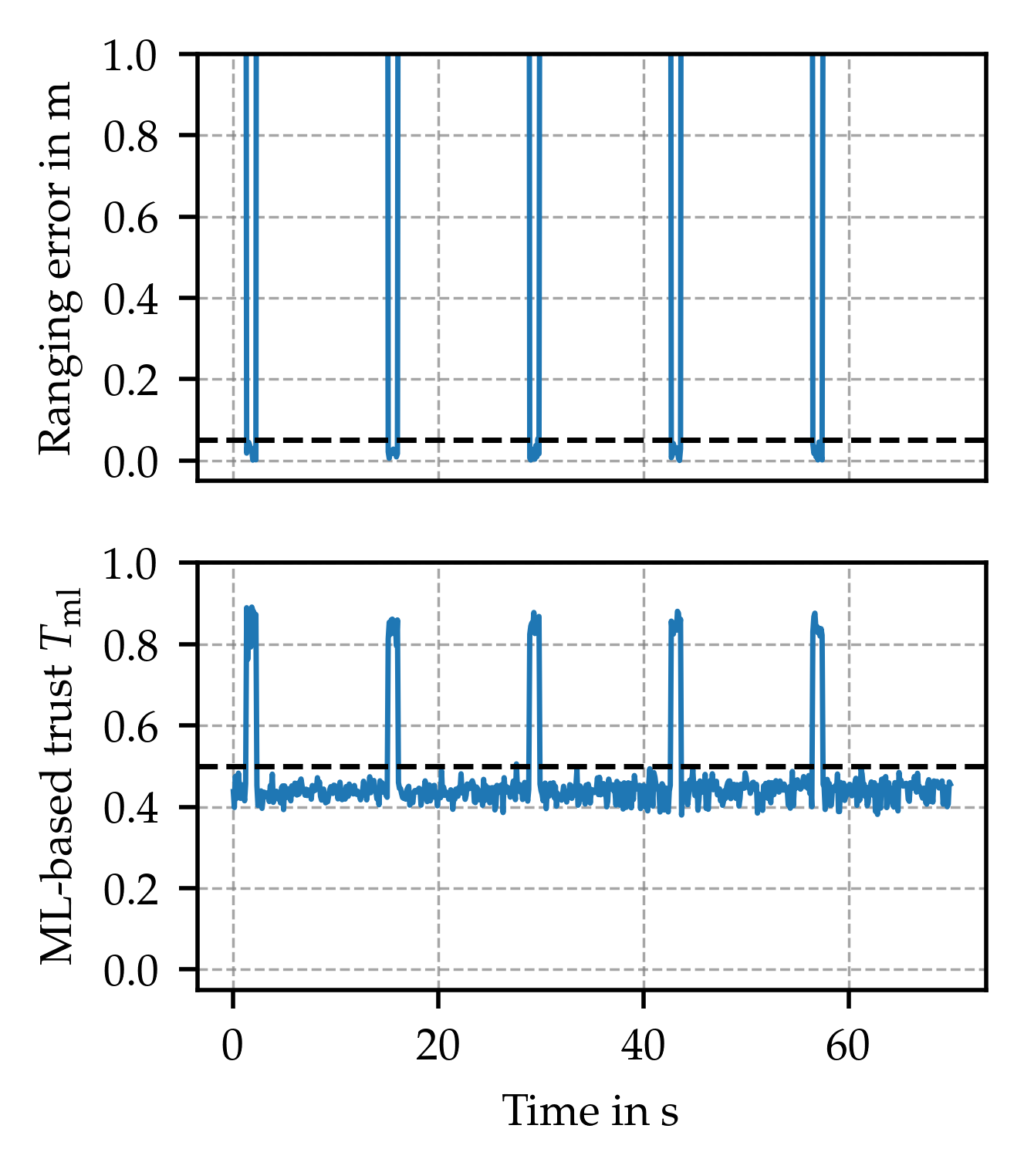}
          \\
         (a) & 
         (b)
    \end{tabular}

    \caption{Performance of ML-based anomaly detection score in scenario (a) channel manipulation by a person temporarily blocking line-of-sight and (b) SHR attack, active approx. 90\% of time.}
    \label{fig:jkuaec-nlos-jamming}
\end{figure}

ML-based anomaly detection facilitates channel impulse responses estimated at the receiver to effectively detect channel obstructions, \cref{fig:jkuaec-nlos-jamming}(a), accounting for localization accuracy. \bl{The method evaluates the distance between key features of the CIR from known trustworthy channels to key features of the current channel realization. Thus, it indicates if the channel behaves as intended.} Figure~\ref{fig:jkuaec-nlos-jamming}(b) reveals that ML-based anomaly detection can also detect spoofing attacks (e.g., SHR attack \cite{peterseil2023}), i.e., active attackers and interferers. The algorithm itself uses a sigmoid function at the output. Hence, additional mapping is not required. \bl{The trustworthiness index is defined as
\begin{equation*}
    T_\mathrm{ml} = m_\mathrm{ml} = f_\mathrm{ml}(\mathbf{\hat{h}}_\mathrm{a}, \mathbf{\hat{h}}_\mathrm{b}, \mathbf{\hat{h}}_\mathrm{c}),
\end{equation*}
with the autoencoder $f_\mathrm{ml}$ as defined, trained and validated in \cite{peterseil2023}.}

\subsection{Received Signal Strength Indicator (\texorpdfstring{$T_\mathrm{rssi}$}{Trssi})}
To identify anchors that may not be responding reliably, thus affecting service availability, the RSSI reported by the UWB transceiver is assigned to $m_\mathrm{rssi}$. 
In \cref{fig:sigmoid-rssi}(b), the packet error rate is plotted over RSSI. The tuning parameter $\underline{m}_\mathrm{rssi}$ was chosen according to an acceptable packet error rate of 1\%. The evaluation of an UWB data set collected in an office environment \cite{ppeterseil2021} showed that the path loss caused by typical obstacles (i.e., people, bookshelves) does not exceed \SI{5}{\decibel}. Thus,
to ensure sufficient signal strength in varying indoor scenarios
$\overline{m}_\mathrm{rssi} =\underline{m}_\mathrm{rssi}+\SI{5}{\decibel}$ is used as trustworthy condition.
Finally, the mapping
parameters are
$\underline{m}_\mathrm{rssi} = \SI{-92}{\decibel}$ and $\overline{m}_\mathrm{rssi} = \SI{-87}{\decibel}$.

\subsection{Position Dilution of Precision (\texorpdfstring{$T_\mathrm{pdop}$}{Tpdop})}
\begin{figure}
    \centering

    \begin{tabular}{@{}c@{}c@{}}
          \includegraphics[scale=1]{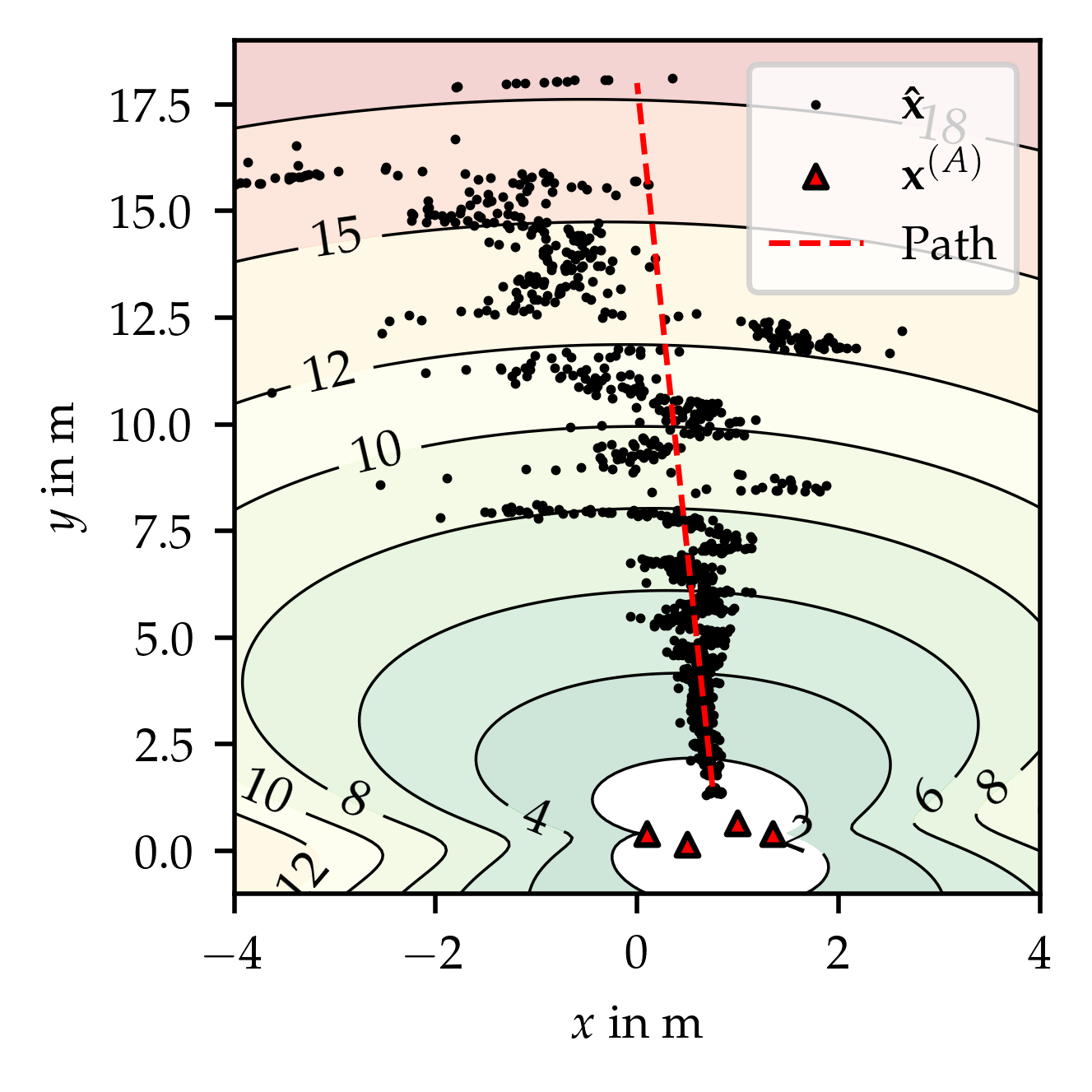}
          &
          \includegraphics[scale=1]{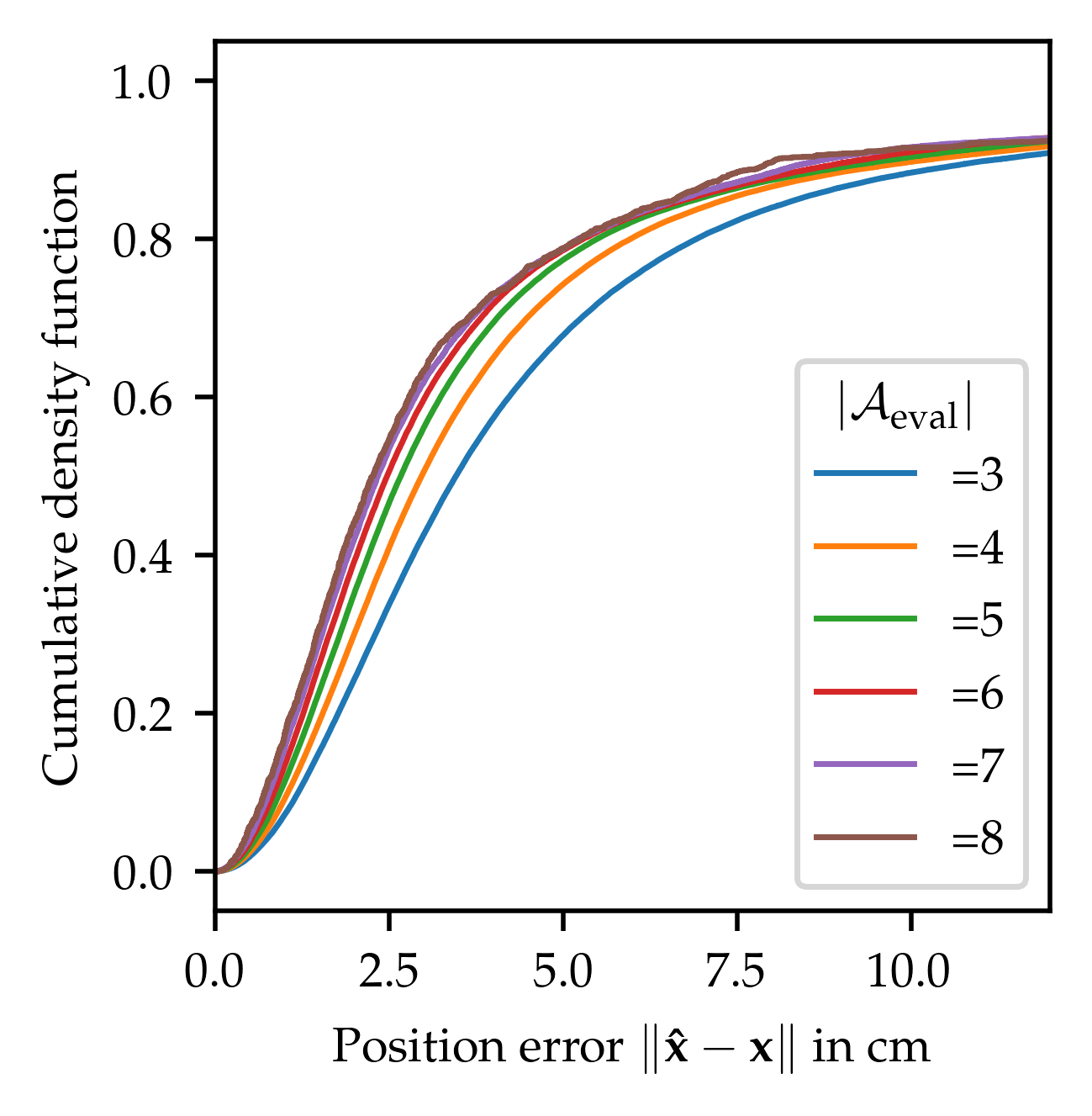}
          \\
         (a) & 
         (b)
    \end{tabular}

    \caption{(a) Localization scenario with improper anchor configuration. The map shows levels of PDoP in the range from $2$ to $18$. (b) Cumulative density function of position error w.r.t the number of anchors used.}
    \label{fig:pdop-nanchors}
\end{figure}

PDoP gives an indication of the accuracy that can be achieved based on the placement of available anchors with respect to the estimated node position. It can be roughly interpreted as a ratio of position error to range error.
\begin{table}
    \centering
    \caption{Rating of PDoP values \cite{robotics9030066}.}
    \label{tab:dop}
    \begin{tabular}{lcccccc}
      \toprule
      $m_\text{pdop}$ & $<$1 & 1–2 & 2–5 & 5–10 & 10–20 & $>$20 \\
      \midrule
      {Rating} & Ideal & Excellent & Good & Moderate & Fair & Poor \\
      \bottomrule
    \end{tabular}
\end{table}
\bl{Remember the anchors used for localization $\mathcal{A}_\text{eval} = \{\text{A}_{1}, \ldots, \text{A}_{K}\}$ and $A \in \mathcal{A}_\mathrm{eval}$. At first the vectors
\begin{equation*}
    \mathbf{c}^{(A)}=\mathbf{x}^{(A)}-\mathbf{\hat{x}},
\end{equation*}
 pointing from the estimated node position $\mathbf{\hat{x}}$, \eqref{eq:localization}, to the respective anchor positions $\mathbf{x}^{(A)}$ are defined. These vectors $\mathbf{c}^{(A)}$ are then normalized and collected in the rows of matrix
\begin{equation*}
    \mathbf{D} = \begin{bmatrix}
        \frac{\mathbf{c}^{(\text{A}_{1})}}{\|\mathbf{c}^{(\text{A}_{1})}\|} ~\text{\scriptsize\rotatebox[origin=c]{90}{$---$}}~ \dots ~\text{\scriptsize\rotatebox[origin=c]{90}{$---$}}~ \frac{\mathbf{c}^{(\text{A}_{K})}}{\|\mathbf{c}^{(\text{A}_{K})}\|}
    \end{bmatrix}^\top.
\end{equation*}
Finally, according to \cite{robotics9030066}, the metric
\begin{equation*}
    m_\mathrm{pdop} = \sqrt{\mathrm{trace}{\left((\mathbf{D}^\top\mathbf{D})^{-1}\right)}}
\end{equation*}
is defined.} Table~\ref{tab:dop} reveals that a good performance of the localization system can be expected for $m_\text{pdop} < 3$, while moderate performance can be achieved up to $m_\text{pdop} < 10$. Figure~\ref{fig:pdop-nanchors}(a) shows position estimates $\mathbf{\hat{x}}$ of a node moving along a linear path collected in an experiment. Based on the experiment, scattering of position estimates significantly increases for $m_\text{pdop} > 8$.
The mapping was defined as
\begin{equation*}
    T_\mathrm{pdop} =
    \begin{cases}
    \zeta\left(m_\mathrm{pdop}, \underline{m}_\mathrm{pdop}, \overline{m}_\mathrm{pdop}\right) & \text{if } |\mathcal{A}_\mathrm{eval}| \geq 3, \\
    1 & \text{otherwise}.
    \end{cases}
\end{equation*}
with tuning parameters $\underline{m}_\mathrm{pdop} = 8$ and $\overline{m}_\mathrm{pdop} = 3$.

\subsection{Number of Anchors (\texorpdfstring{$T_\mathrm{na}$}{Tna})}
The number of anchors in reach,
\begin{equation*}
    m_{na} = |\mathcal{A}_\mathrm{eval}|,
\end{equation*}
gives a measure of redundancy. At least three distance estimates, i.e., available anchors, are required for 2D localization. Having more anchors available increases the service's fault tolerance. Figure~\ref{fig:pdop-nanchors}(b) shows that a higher number of anchors can also improve the accuracy to a limited extent.
The mapping was chosen
with $\underline{m}_\mathrm{na} = 3$ and $\overline{m}_\mathrm{na} = 4.5$.

\subsection{Binary Trustworthiness Indicators (\texorpdfstring{$T_\mathrm{auth}$}{Tauth}, \texorpdfstring{$T_\mathrm{enc}$}{Tenc}, \texorpdfstring{$T_\mathrm{sr}$}{Tsr} and \texorpdfstring{$T_\mathrm{da}$}{Tda})}

Binary metrics measure the system state by checking if authentication, encryption, secure ranging, or dynamic addressing is used. Authentication and encryption are implemented using authenticated encryption with associated data according to the IEEE802.15.4 standard, ensuring data integrity and confidentiality. Secure ranging, an enhancement using the STS option from the IEEE802.15.4z standard, partially protects against several physical layer attacks. Dynamic addressing involves nodes and anchors changing their identifiers pseudorandomly after each ranging cycle, increasing privacy by obfuscating the identities of communicating devices.
As the use of each scheme adds extra protection, the mapping to the binary trustworthiness indicator $T'$ is
\begin{equation*}
    T'=
    \begin{cases}
    1 & \text{if the corresponding scheme is used,} \\
    0 & \text{otherwise}.
    \end{cases}
\end{equation*}
Note that even if these indicators reflect system settings, it is imperative to evaluate them continuously, as sophisticated attacks may alter these settings.

\section{Evaluation}

\begin{figure}
    \centering

    \includegraphics[]{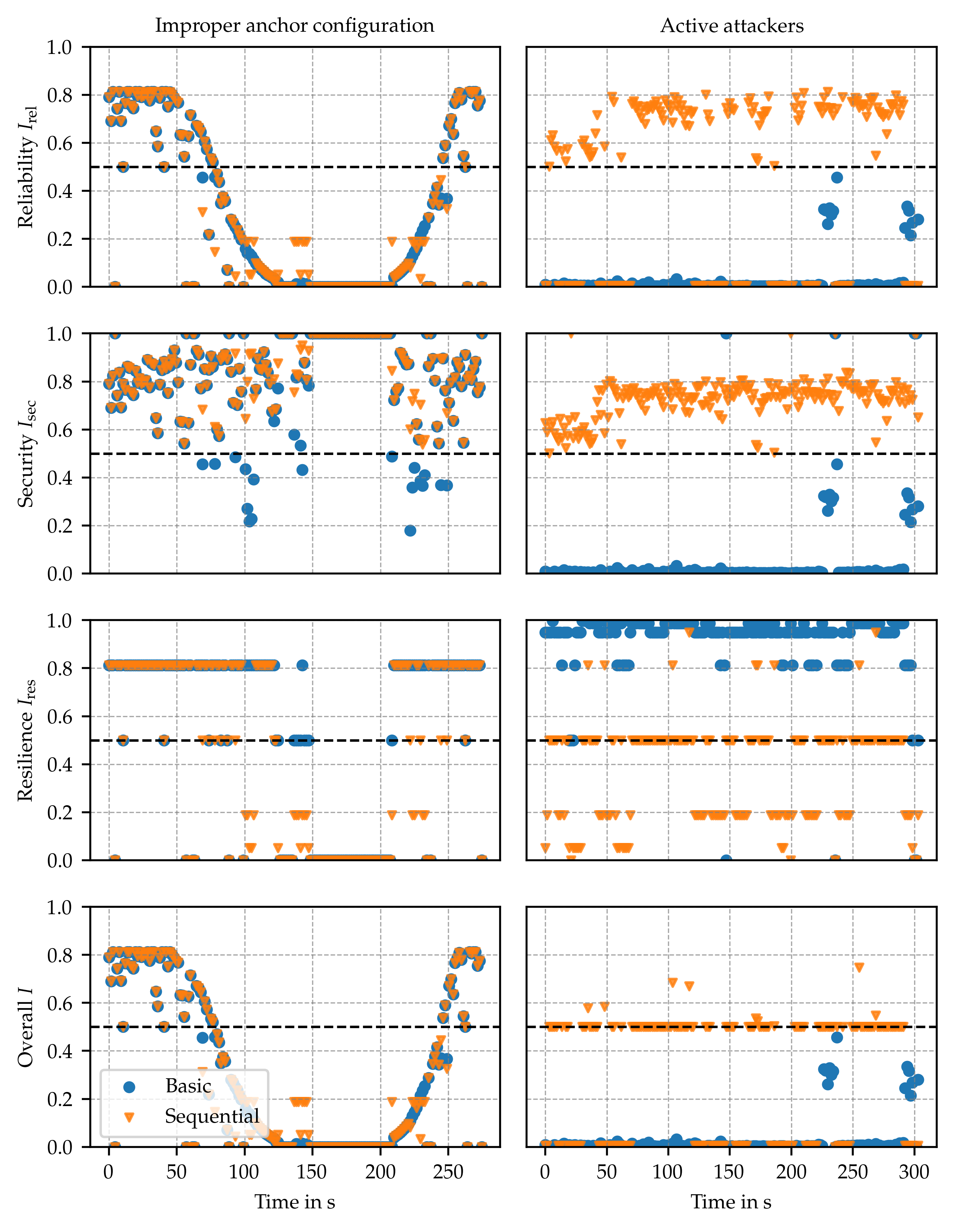}

    \caption{Trustworthiness evaluation on two threats. For the threat of improper anchor configuration, the distance from the node to a set of anchors is increased (left column). For the threat of active attackers, 4 out of 8 anchors were subject to SHR jamming (right column).}
    \label{fig:eval-indices}
\end{figure}

\begin{figure}
    \centering

    \includegraphics[]{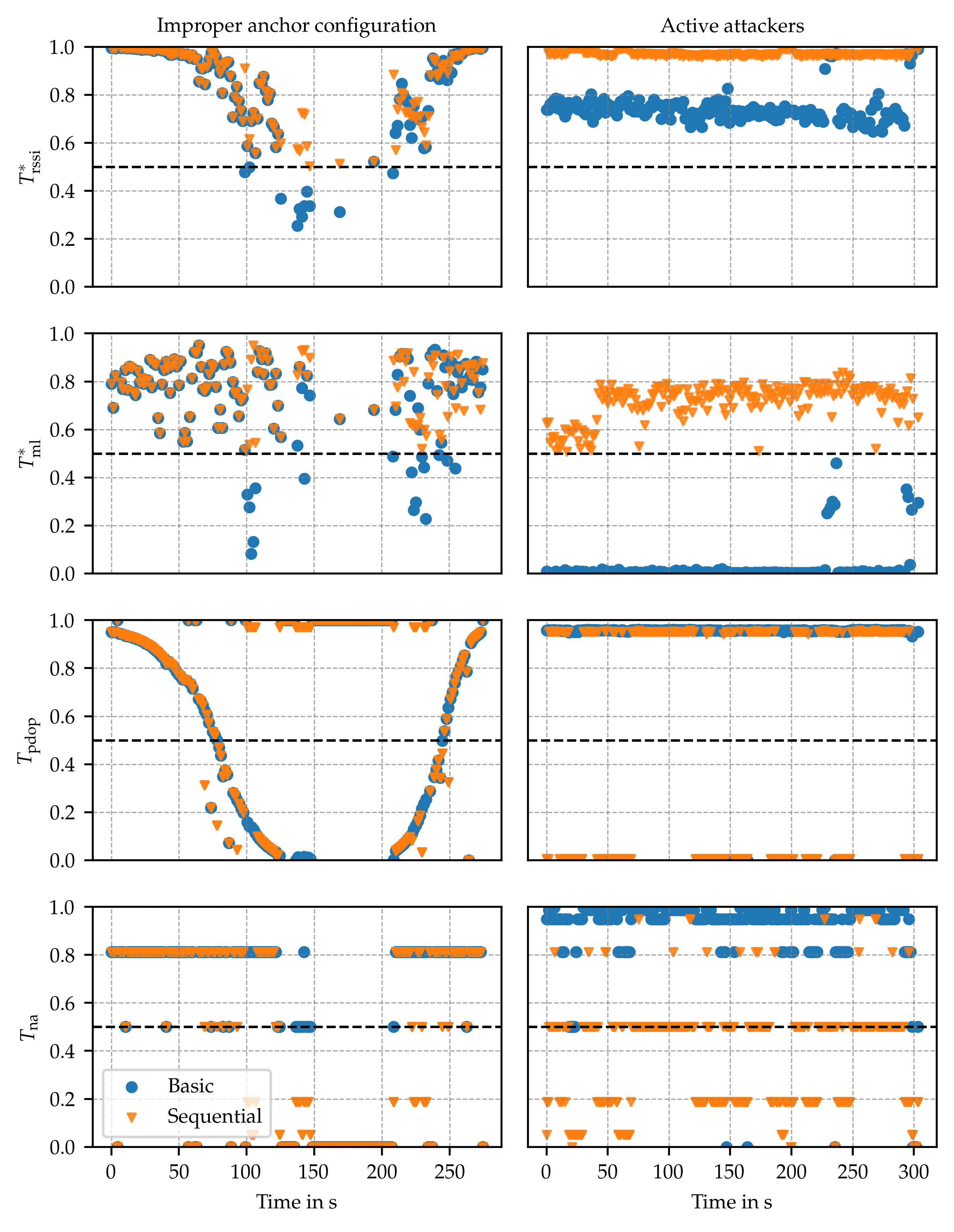}

    \caption{Trustworthiness indicators during evaluation scenario \textit{Improper anchor configuration} (left) and \textit{active attackers} (right). For depiction, a subset of indicators is selected, for such significant changes can be observed. These are the combined link indicators $T^\ast_\text{rssi}$ and ML-based anomaly detection $T^\ast_\text{ml}$ as well as the system indicators on PDoP $T_\text{pdop}$ and number of anchors $T_\text{na}$.}
    \label{fig:eval-indicators}
\end{figure}

\bl{The trustworthiness assessment is designed to reflect the system's operational trustworthiness across various attributes, including reliability, security, privacy, resilience, and safety. This assessment is not tailored to address specific threats. Instead, a threat analysis, similar to sensitivity analysis, is conducted to identify relevant parameters. Ideally, this process encompasses all significant observation parameters that indicate proper system behavior. Consequently, the trustworthiness assessment must be capable of signaling low trustworthiness in the presence of any threat identified in \cref{sec:threat-analysis}, as well as detecting unknown system issues that impact these parameters. In this evaluation, the influence of two specific threats on the trustworthiness assessment is demonstrated, }
namely, improper anchor configuration
and active attackers (c.f. \cref{sec:threat-analysis}). 
\bl{One and the same implementation of the trustworthiness framework was used for both scenarios, without any tuning to the specific conditions of each scenario. The machine learning-based anomaly detection indicator was trained using a dataset \cite{ppeterseil2021} comprising range estimates captured in line-of-sight conditions across different environments (an auditorium and a private workshop) than those used in this work.}
The results are summarized in \cref{fig:eval-indices}, and in \cref{fig:eval-indicators}, where, respectively, trustworthiness indices ($I_\text{rel}, I_\text{sec}, I_\text{res}$ and $I$ ) and trustworthiness indicators ($T_\text{rssi}^{*}, T_\text{ml}^{*}, T_\text{pdop}$ and $T_\text{na}$), significant for the selected scenarios, are depicted.
In both scenarios, the assessment methods (c.f. \cref{sec:assessment-anchor-selection}) basic (colored in blue) and sequential (colored in orange) are compared against each other.




\subsection{Improper Anchor Configuration}\label{sec:eval_inproper_anchor_config}
In this experiment, corresponding to the setup used in \cref{fig:pdop-nanchors}(a), the node first approaches the borders of the service area covered by a set of four anchors, exceeds them, and then, at a maximum distance of approximately 20 meters from the anchors, returns to its starting position.
Initially, the anchors are favorably located, i.e. ($T_\text{pdop}$) is low and the signal strengths ($T^{*}_\text{rssi}$) are high, resulting in a high reliability index ($I_\mathrm{rel}$) dominated by the number of anchors available ($T_\text{na}$). As the distance to the anchors increases, the dilution of precision $T_\text{pdop}$ and signal strength $T^{*}_\text{rssi}$ decrease. After 50 seconds, $T_\text{pdop}$ starts dominating the reliability index, causing $I_\text{rel}$ to decline. This decline in trustworthiness corresponds to a loss in accuracy due to the dilution of precision effect. After approximately 75 seconds, the index reaches the threshold, signaling the transition to an untrustworthy state. As the node moves further away from the anchors, the signal strength $T^{*}_\text{rssi}$ after 100 seconds also reaches the threshold, i.e., when reduced service availability is anticipated. Breaking links reduce the number of anchors $T_\text{na}$ and cause the system to fail.

With this experiment, the framework's ability to classify the system's state as untrustworthy well before a complete loss of localization service occurs is demonstrated. This early detection provides an opportunity to implement countermeasures in a timely manner.
Recall that the sequential assessment method differs from the basic method by selecting the subset of anchors with trustworthy link indicators (c.f. \cref{sec:assessment-anchor-selection}). In this experiment, both assessment methods performed similarly. This is due to the fact that $T_{\text{ml}^{(A)}}$ for all $A \in \mathcal{A}$ stayed at a high level as well as $T_{\text{rssi}^{(A)}}$ for all $A \in \mathcal{A}$ were at similar levels at any point in time.


\subsection{Active attackers}
In this scenario, eight anchors are used in a static office environment, positioned at distances ranging from \SI{1.5}{\meter} to \SI{4}{\meter} to the node. A jamming device in proximity of the node is executing an SHR attack on four out of eight anchors, aiming at the reduction of estimated distances, which further yields wrong position estimates.
Using the basic method, $I_\mathrm{rel}$ and $I_\mathrm{sec}$ are both low due to low values $T_\mathrm{ml}^{*}$ for anchors being subject to jamming. However, $I_\mathrm{res}$ is high due to a larger anchor set $\mathcal{A}_\text{eval}$, also reflected in $T_\text{na}$.

Using the sequential method, in the first step, the trustworthiness of anchor links is evaluated. Based on the subset of trustworthy anchors, in the second step, the location estimate, the remaining trustworthiness indicators and indices are computed. In $I_\mathrm{sec}$, this is reflected by maintaining trustworthiness at a level of approximately 0.6, originating from $T_\mathrm{ml}$ of trusted anchors, i.e., from anchors not subject to jamming. However, by considering a lower number of anchors for the evaluation, $I_\mathrm{res}$ is around 0.5, and for some measurements, it is even lower. 
Despite the lower resilience value, the overall trustworthiness index $I$ indicates that the sequential method can maintain tolerable trustworthiness during the attack on 4 out of 8 anchors.

This benefit can also be seen in the accuracy. While the root mean square error in the attack scenario would result in approx. \SI{81}{\centi\meter}, the basic method classifies all estimates as untrustworthy. The sequential method achieves a root mean square error of \SI{17}{\centi\meter} with a trustworthiness index $I \geq 0.5$. In only 39\% of the estimates it obtains $I<0.5$.

\subsection{Findings}

This subsection summarizes the essential findings from the experimental evaluation.

\begin{itemize}
    \item In both examples, $I_\mathrm{rel}$ and $I_\mathrm{sec}$ precisely detect threats to availability, accuracy, and integrity. Since reliability and security are attributes that support safety, this also indicates that the proposed method can detect certain safety threats.
    \item The improved reliability $I_\text{rel}$ and security $I_\text{sec}$ of the \textit{sequential} method in the active attacker use case underlines the potential of using intermediate results from the trustworthiness assessment to increase the service performance.
    \item To provide a holistic prediction of the system's proper operation, trustworthiness assessment through carefully selected metrics is essential. Several of these metrics influence multiple attributes, resulting in an interconnected evaluation that advances beyond the isolated analysis of each attribute.
    \item From changes within the trustworthiness indices, one may predict system vulnerabilities before the actual occurrence of failures, c.f., $I_\text{rel}$ in \cref{sec:eval_inproper_anchor_config}. Hence, trustworthiness has a high potential to be leveraged as an early warning mechanism. Furthermore, this early warning offers the possibility of taking countermeasures to maintain the level of trustworthiness in the system.
\end{itemize}




\section{Conclusion}
In this work, a method that systematically links the general definition of trustworthiness to an evidence-based trustworthiness index is proposed. The focus is on UWB self-localization, a critical service in the IoT domain. The threat-driven metric selection approach represents the first holistic assessment of trustworthiness concerning reliability, security, privacy, and resilience. While safety is often seen as an additional attribute, in the context of UWB self-localization, its characteristics are considered to be supported by reliability and security.

The proposed method, which connects trustworthiness definitions and attributes to threats, metrics, and trustworthiness indicators and indices, has the potential to serve as a general framework. While future work may extend the threat analysis and metric selection, the proposed approach demonstrates the functional principle. Interestingly, the interconnection of attributes through individual metrics indicates that a holistic evaluation of trustworthiness surpasses the isolated analysis of individual attributes.

Experimental analysis shows that using intermediate trustworthiness indicators can improve service quality. Comparing conventional UWB self-localization with an enhanced method that applies a trustworthiness-based anchor selection scheme reveals clear localization performance improvements without additional costs. Furthermore, while many traditional metrics are model-based or derived from information theory, ML techniques can also significantly contribute to the assessment of trustworthiness.

In conclusion, the presented method proposes a systematic approach for holistic trustworthiness assessment. By leveraging intermediate results and incorporating advanced techniques such as ML-based metrics, substantial improvements in system performance can be achieved, highlighting the potential for future advancements.

\vspace{6pt}


\authorcontributions{Conceptualization, P.P., B.E., J.H. and R.K.; methodology, P.P., B.E., J.H. and R.K.; software, P.P.; validation, P.P. and A.S.; formal analysis, P.P.; investigation, B.E., R.K.; resources, P.P. and B.E.; data curation, P.P.; writing---original draft preparation, P.P., B.E., J.H. and R.K.; writing---review and editing, P.P., B.E., J.H., R.K. and A.S.; visualization, P.P., B.E., J.H. and R.K.; supervision, A.S.; project administration, B.E. and P.P.; funding acquisition, A.S. All authors have read and agreed to the published version of the manuscript.}


\funding{This work has been carried outwithin the scope of Digidow, the Christian Doppler Laboratoryfor Private Digital Authentication in the Physical World and haspartially been supported by the LIT Secure and Correct SystemsLab, with financial support by the Austrian Federal Ministry of Labour and Economy and the State of Upper Austria.
}

\conflictsofinterest{The authors declare no conflicts of interest.}

\begin{adjustwidth}{-\extralength}{0cm}

\reftitle{References}
\bibliography{main}

\end{adjustwidth}
\end{document}

%% file: texfig/DSTWR_CIR.tex
\begin{tikzpicture}[node distance = 0pt]

\tikzset{>={latex[length=2mm]}}

\def\x{2.2cm}
\def\height{1.0cm} 
\def\offset{0.9cm} 
\def\Rt{1.2cm}
\def\Da{0.8cm}
\def\Ra{1.2cm}
\def\Dt{0.8cm}
\def\Rtt{1.2cm}
\def\Dat{0.8cm}

\def\labeldist{0.5cm}

\pgfmathsetmacro\Tof{0.5*\Rt-0.5*\Da}
\def\bsep{4.6cm}

\node[above,minimum height = 2em] (T0) at (0,0) {\footnotesize Node};
\node[] (T1) at (0,-\height) {};
\node[above,minimum height = 2em] (A0) at ($(T0.south)+(\x,0)$) {\footnotesize $\text{A}_k$};
\node[] (A1) at (\x,-\height) {};
\node[] (TXb) at ($(A0.south)+(0,0.5*\Tof-0.2*\height-\offset)$){};
\node[] (RXb) at ($(T0.south)+(0,-0.5*\Tof-0.2*\height-\offset)$){};
\node[] (TXa) at ($(RXb) + (0,\Rt)$){};
\node[] (RXa) at ($(TXb)+(0,\Da)$){};
\node[] (TXc) at ($(RXb)-(0,\Dt)$){};
\node[] (RXc) at ($(TXb)-(0,\Ra)$){};
\node[] (TXd) at ($(RXc)-(0,\Dat)$){};
\node[] (RXd) at ($(TXc)-(0,\Rtt)$){};

\node[label=left :{\scriptsize {$t_\mathrm{a}$}}]                                       (TXalabel) at ($(TXa)-(\labeldist,0)$){};
\node[label=right:{\scriptsize {$\tau_\mathrm{a}, \mathbf{\hat{h}}_\mathrm{a}$}}]                (RXalabel) at ($(RXa)+(\labeldist,0)$){};
\node[label=right :{\scriptsize {$\tau_\mathrm{b}$}}]                                      (TXblabel) at ($(TXb)+(\labeldist,0)$){};
\node[label=left :{\scriptsize {$\mathrm{RSSI}, \mathbf{\hat{h}}_\mathrm{b}, t_\mathrm{b}$}}] (RXblabel) at ($(RXb)-(\labeldist,0)$){};
\node[label=left :{\scriptsize {$t_\mathrm{c}$}}]                                       (TXclabel) at ($(TXc)-(\labeldist,0)$){};
\node[label=right:{\scriptsize {$\tau_\mathrm{c}, \mathbf{\hat{h}}_\mathrm{c}$}}]                (RXclabel) at ($(RXc)+(\labeldist,0)$){};

\draw[->,dashed] (TXa.center) -- (RXa.center) node[pos = 0.5, sloped,above] {\footnotesize Packet a};
\draw[->,dashed] (TXb.center) -- (RXb.center) node[pos = 0.5, sloped,above] {\footnotesize Packet b};
\draw[->,dashed] (TXc.center) -- (RXc.center) node[pos = 0.5, sloped,above] {\footnotesize Packet c};
\draw[->,dashed] (TXd.center) -- (RXd.center) node[pos = 0.5, sloped,above] {\footnotesize Packet d};

\def\sep{0.2}
\node[left = \sep cm of TXa.center] (Rt1) {};
\node[left = \sep cm of RXb.center] (Rt2) {};
\node[right = \sep cm of RXa.center] (Da1) {};
\node[right = \sep cm of TXb.center] (Da2) {};
\node[right = \sep cm of TXb.center] (Ra1) {};
\node[right = \sep cm of RXc.center] (Ra2) {};
\node[left = \sep cm of RXb.center] (Dt1) {};
\node[left = \sep cm of TXc.center] (Dt2) {};

\draw[|<->] (Rt1.center) -- (Rt2.center) node[pos=.5,left] {\scriptsize$R_n$};
\draw[|<->] (Da1.center) -- (Da2.center) node[pos=.5,right] {\scriptsize$D_a$};
\draw[|<->|] (Ra1.center) -- (Ra2.center) node[pos=.5,right] {\scriptsize$R_a$};
\draw[|<->|] (Dt1.center) -- (Dt2.center) node[pos=.55,left] {\scriptsize$D_n$};

\draw[{-Stealth}] ($(TXa.center)+ (0,0.1cm)$) -- ($(TXa.center) + (0,-4cm)$) node[left] {\footnotesize Time};

\coordinate (CA) at (0,0);
\begin{axis}[
x axis line style={draw=none},
y axis line style={draw=none},
height=1cm,
axis lines = middle,
scale only axis=true,
width= \height,
ytick=\empty,
xmin=0.7,xmax=1.35,
xtick distance=1,
hide obscured x ticks=false,
xlabel style={rotate=90,anchor = south west, fill=white},
xticklabel style={anchor = east},
xtick=\empty,
ymin=0, ymax=16000.1,
y dir=reverse,
at = {($(TXa.center) + (0,-1.05cm)$)},
rotate around={-90:($(current axis.origin)-(CA)$)},
anchor = origin,
]
\addplot [thin, color1]
table [row sep=\\] {%
0.73143 161\\0.74012 95\\0.74882 196\\0.75752 137\\0.76621 57\\0.77491 117\\0.7836 290\\0.7923 87\\0.80099 76\\0.80969 282\\0.81839 1618\\0.82708 7730\\0.83578 8342\\0.84447 5975\\0.85317 5790\\0.86186 8075\\0.87056 3765\\0.87925 3358\\0.88795 2059\\0.89665 246\\0.90534 1769\\0.91404 3944\\0.92273 1956\\0.93143 265\\0.94012 1881\\0.94882 969\\0.95752 64\\0.96621 1004\\0.97491 854\\0.9836 330\\0.9923 789\\1.001 683\\1.0097 1277\\1.0184 515\\1.0271 708\\1.0358 1206\\1.0445 332\\1.0532 519\\1.0619 454\\1.0706 241\\1.0793 331\\1.088 938\\1.0966 1055\\1.1053 565\\1.114 1164\\1.1227 949\\1.1314 375\\1.1401 370\\1.1488 694\\1.1575 542\\1.1662 423\\1.1749 341\\1.1836 33\\1.1923 562\\1.201 782\\1.2097 352\\1.2184 147\\1.2271 259\\1.2358 145\\1.2445 74\\1.2532 125\\1.2619 295\\1.2706 338\\1.2793 258\\1.288 207\\1.2966 242\\1.3053 89\\1.314 86\\1.3227 59\\1.3314 281\\};
\end{axis}

\coordinate (mid) at ($0.5*(RXa.center)+0.5*(TXb.center)$);

\draw[->] ($(TXa.center)+ (\x,0.1cm)$) -- ($(TXa.center) + (\x,-4cm)$) node[left] {};
\begin{axis}[
x axis line style={draw=none}, 
y axis line style={draw=none},
height=1cm,
axis lines = middle,
scale only axis=true,
width=\height,
ytick=\empty,
xmin=-0.25,xmax=0.4,
xtick distance=1,
hide obscured x ticks=false,
xtick=\empty,
xlabel style={rotate=90,anchor = south west},
xticklabel style={anchor = east, fill=white, fill opacity=0.7, text opacity=1},
ymin=0, ymax=16000.1,
at = {($(mid) + (0,0.15cm)$)},
rotate around={-90:($(current axis.origin)-(CA)$)},
anchor = origin,
]
\addplot [thin, color0]
table [row sep=\\] {%
-0.21857 142\\-0.20988 144\\-0.20118 167\\-0.19248 43\\-0.18379 118\\-0.17509 135\\-0.1664 177\\-0.1577 96\\-0.14901 123\\-0.14031 306\\-0.13161 2034\\-0.12292 7779\\-0.11422 7483\\-0.10553 3421\\-0.096832 7363\\-0.088137 7475\\-0.079441 3599\\-0.070745 2776\\-0.06205 1551\\-0.053354 712\\-0.044658 3360\\-0.035963 4909\\-0.027267 740\\-0.018571 441\\-0.0098758 2153\\-0.0011801 867\\0.0075155 337\\0.016211 1143\\0.024907 847\\0.033602 556\\0.042298 725\\0.050994 851\\0.059689 1261\\0.068385 266\\0.077081 1323\\0.085776 1166\\0.094472 65\\0.10317 473\\0.11186 430\\0.12056 327\\0.12925 518\\0.13795 1183\\0.14665 1100\\0.15534 837\\0.16404 1392\\0.17273 457\\0.18143 421\\0.19012 453\\0.19882 740\\0.20752 499\\0.21621 714\\0.22491 347\\0.2336 109\\0.2423 509\\0.25099 854\\0.25969 383\\0.26839 46\\0.27708 275\\0.28578 119\\0.29447 68\\0.30317 91\\0.31186 360\\0.32056 347\\0.32925 266\\0.33795 121\\0.34665 155\\0.35534 149\\0.36404 235\\0.37273 33\\0.38143 167\\};
\end{axis}

\begin{axis}[
x axis line style={draw=none}, 
y axis line style={draw=none},
height=1cm,
axis lines = middle,
scale only axis=true,
width=\height,
ytick=\empty,
xmin=-0.25,xmax=0.4,
xtick distance=1,
hide obscured x ticks=false,
xtick=\empty,
xlabel style={rotate=90,anchor = south west},
xticklabel style={anchor = east, fill=white, fill opacity=0.7, text opacity=1},
ymin=0, ymax=16000.1,
at = {($(mid) + (0,-1.85cm)$)},
rotate around={-90:($(current axis.origin)-(CA)$)},
anchor = origin,
]
\addplot [thin, color0]
table [row sep=\\] {%
-0.21857 142\\-0.20988 144\\-0.20118 167\\-0.19248 43\\-0.18379 118\\-0.17509 135\\-0.1664 177\\-0.1577 96\\-0.14901 123\\-0.14031 306\\-0.13161 2034\\-0.12292 7779\\-0.11422 7483\\-0.10553 3421\\-0.096832 7363\\-0.088137 7475\\-0.079441 3599\\-0.070745 2776\\-0.06205 1551\\-0.053354 712\\-0.044658 3360\\-0.035963 4909\\-0.027267 740\\-0.018571 441\\-0.0098758 2153\\-0.0011801 867\\0.0075155 337\\0.016211 1143\\0.024907 847\\0.033602 556\\0.042298 725\\0.050994 851\\0.059689 1261\\0.068385 266\\0.077081 1323\\0.085776 1166\\0.094472 65\\0.10317 473\\0.11186 430\\0.12056 327\\0.12925 518\\0.13795 1183\\0.14665 1100\\0.15534 837\\0.16404 1392\\0.17273 457\\0.18143 421\\0.19012 453\\0.19882 740\\0.20752 499\\0.21621 714\\0.22491 347\\0.2336 109\\0.2423 509\\0.25099 854\\0.25969 383\\0.26839 46\\0.27708 275\\0.28578 119\\0.29447 68\\0.30317 91\\0.31186 360\\0.32056 347\\0.32925 266\\0.33795 121\\0.34665 155\\0.35534 149\\0.36404 235\\0.37273 33\\0.38143 167\\};
\end{axis}

\end{tikzpicture}

%% file: texfig/packet_structure.tex
\begin{tikzpicture}[
 lbox/.style={minimum height=6mm, minimum width=18mm, draw=black},
 sbox/.style={minimum height=6mm, minimum width=8mm, draw=black}
]

    \begin{scope}
        \node[lbox,rounded rectangle,rounded rectangle east arc=none] (a) at (0,0) {\footnotesize SYNC};
        \node[anchor=west, sbox] (b) at (a.east) {\footnotesize SFD};
        \draw[-stealth,red!50!black] (b.east) -- ($(b.east) + (0,5mm)$);
        \node[anchor=west, sbox] (c) at (b.east) {\footnotesize PHR};
        \node[anchor=west, lbox,rounded rectangle,rounded rectangle west arc=none] (d) at (c.east) {\footnotesize PHY Payload};

        \node[align=right] at (6,0) {\footnotesize Config: 0};
    \end{scope}

    \begin{scope}[yshift=-10mm]
        \node[lbox,rounded rectangle,rounded rectangle east arc=none] (a) at (0,0) {\footnotesize SYNC};
        \node[anchor=west, sbox] (b) at (a.east) {\footnotesize SFD};
        \draw[-stealth,red!50!black] (b.east) -- ($(b.east) + (0,5mm)$);
        \node[anchor=west, lbox] (c) at (b.east) {\footnotesize STS};
        \node[anchor=west, sbox] (d) at (c.east) {\footnotesize PHR};
        \node[anchor=west, lbox,rounded rectangle,rounded rectangle west arc=none] (e) at (d.east) {\footnotesize PHY Payload};
        
        \node[align=right] at (6.46,0) {\footnotesize 1};
    \end{scope}

    \begin{scope}[yshift=-20mm]
        \node[lbox,rounded rectangle,rounded rectangle east arc=none] (a) at (0,0) {\footnotesize SYNC};
        \node[anchor=west, sbox] (b) at (a.east) {\footnotesize SFD};
        \draw[-stealth,red!50!black] (b.east) -- ($(b.east) + (0,5mm)$);
        \node[anchor=west, sbox] (c) at (b.east) {\footnotesize PHR};
        \node[anchor=west, lbox] (d) at (c.east) {\footnotesize PHY Payload};
        \node[anchor=west, lbox,rounded rectangle,rounded rectangle west arc=none] (e) at (d.east) {\footnotesize STS};

        \node[align=right] at (6.46,0) {\footnotesize 2};
    \end{scope}

    \begin{scope}[yshift=-30mm]
        \node[lbox,rounded rectangle,rounded rectangle east arc=none] (a) at (0,0) {\footnotesize SYNC};
        \node[anchor=west, sbox] (b) at (a.east) {\footnotesize SFD};
        \draw[-stealth,red!50!black] (b.east) -- ($(b.east) + (0,5mm)$);
        \node[anchor=west, lbox,rounded rectangle,rounded rectangle west arc=none] (c) at (b.east) {\footnotesize STS};

        \node[text width=2.4cm,execute at begin node=\setlength{\baselineskip}{1.5ex}] at (4.7,0) {\scriptsize Arrows indicate timestamping event};

        \node[align=right] at (6.46,0) {\footnotesize 3};
    \end{scope}    
\end{tikzpicture}

%% file: main.bbl
\begin{thebibliography}{999}

\bibitem[Feng(2017)]{feng2017trusted}
Feng, D.
\newblock {\em Trusted Computing: Principles and Applications}; Vol.~2, Walter
  de Gruyter GmbH \& Co KG,  2017.

\bibitem[{(CPS PWG)}(2016)]{NIST2016cpsframework}
{(CPS PWG)}, C.P.S.P.W.G.
\newblock Framework for Cyber-Physical Systems.
\newblock White Paper Release 1.0, National Institute of Standards and
  Technology {(NIST)},  2016.

\bibitem[Buchheit et~al.()Buchheit, Hirsch, and Martin]{buchheit30industrial}
Buchheit, M.; Hirsch, F.; Martin, R.A.
\newblock The Industrial Internet of Things trustworthiness framework
  foundations.
\newblock {\em Industrial Internet Consortium. https://www. iiconsortium.
  org/pdf/Trustworthi ness\_Framework\_Foundations. pdf (accessed Dec. 30,
  2021)}.

\bibitem[ITU-T(2017)]{ITUT2017Overview}
ITU-T.
\newblock Overview of trust provisioning in information and communication
  technology infrastructures and services.
\newblock ITU-T Recommendation Y.3052, International Telecommunication Union,
  Telecommunication Standardization Sector (ITU-T),  2017.

\bibitem[Cho et~al.(2019)Cho, Xu, Hurley, Mackay, Benjamin, and
  Beaumont]{cho2019stram}
Cho, J.H.; Xu, S.; Hurley, P.M.; Mackay, M.; Benjamin, T.; Beaumont, M.
\newblock Stram: Measuring the trustworthiness of computer-based systems.
\newblock {\em ACM Comp. Surv.} {\bf 2019}, {\em 51},~1--47.

\bibitem[Li et~al.(2020)Li, Zhuang, Hu, Gao, Hu, Chen, He, Pei, Chen, Wang,
  et~al.]{li2020toward}
Li, Y.; Zhuang, Y.; Hu, X.; Gao, Z.; Hu, J.; Chen, L.; He, Z.; Pei, L.; Chen,
  K.; Wang, M.;  et~al.
\newblock Toward location-enabled {IoT} {(LE-IoT)}: {IoT} positioning
  techniques, error sources, and error mitigation.
\newblock {\em IEEE Internet of Things Journal} {\bf 2020}, {\em
  8},~4035--4062.

\bibitem[Coppens et~al.(2022)Coppens, De~Poorter, Shahid, Lemey,
  Van~Herbruggen, and Marshall]{coppens2022overviewUWB}
Coppens, D.; De~Poorter, E.; Shahid, A.; Lemey, S.; Van~Herbruggen, B.;
  Marshall, C.
\newblock An Overview of {UWB} Standards and Organizations ({IEEE} 802.15. 4,
  {FiRa}, Apple): {I}nteroperability Aspects and Future Research Directions.
\newblock {\em IEEE Access} {\bf 2022}, pp. 1--23.

\bibitem[Durand et~al.(2019)]{IIOT2019trustworthiness}
Durand, J.;  et~al.
\newblock The Industrial Internet of Things: Managing and Assessing
  Trustworthiness for {IIoT} in Practice.
\newblock White Paper v1.0, Industrial Internet Consortium, Object Management
  Group, Inc.,  2019.

\bibitem[Xu et~al.(2011)Xu, Gao, Wan, and Xiong]{xu2011trust}
Xu, X.; Gao, X.; Wan, J.; Xiong, N.
\newblock Trust index based fault tolerant multiple event localization
  algorithm for WSNs.
\newblock {\em Sensors} {\bf 2011}, {\em 11},~6555--6574.

\bibitem[Kim et~al.(2019)Kim, Goyat, Rai, Kumar, Buchanan, Saha, and
  Thomas]{kim2019novel}
Kim, T.H.; Goyat, R.; Rai, M.K.; Kumar, G.; Buchanan, W.J.; Saha, R.; Thomas,
  R.
\newblock A novel trust evaluation process for secure localization using a
  decentralized blockchain in wireless sensor networks.
\newblock {\em IEEE access} {\bf 2019}, {\em 7},~184133--184144.

\bibitem[Jain et~al.(2021)Jain, Schott, Scheithauer, H{\"a}ring, H{\"o}flinger,
  Fischer, Habets, Gelhausen, Schindelhauer, and Rupitsch]{jain2021simulation}
Jain, A.K.; Schott, D.J.; Scheithauer, H.; H{\"a}ring, I.; H{\"o}flinger, F.;
  Fischer, G.; Habets, E.A.; Gelhausen, P.; Schindelhauer, C.; Rupitsch, S.J.
\newblock Simulation-Based Resilience Quantification of an Indoor Ultrasound
  Localization System in the Presence of Disruptions.
\newblock {\em Sensors} {\bf 2021}, {\em 21},~6332.

\bibitem[Peterseil et~al.(2022)Peterseil, Etzlinger, M\"arzinger, Khanzadeh,
  and Springer]{peterseil2022datatrust}
Peterseil, P.; Etzlinger, B.; M\"arzinger, D.; Khanzadeh, R.; Springer, A.
\newblock Data Trustworthiness for {UWB} Ranging in {IoT}.
\newblock In Proceedings of the Global Commun. Conf. IEEE,  Dec. 2022, pp.
  1--6.
\newblock (to be presented).

\bibitem[Neirynck et~al.(2016)Neirynck, Luk, and
  McLaughlin]{neirynck2016alternative}
Neirynck, D.; Luk, E.; McLaughlin, M.
\newblock An alternative double-sided two-way ranging method.
\newblock In Proceedings of the Workshop on Pos., Nav. and Commun. (WPNC).
  IEEE,  2016, pp. 1--4.

\bibitem[Cheng et~al.(2012)Cheng, Wu, Zhang, Wu, Li, and
  Maple]{cheng2012localizationsurvey}
Cheng, L.; Wu, C.; Zhang, Y.; Wu, H.; Li, M.; Maple, C.
\newblock A survey of localization in wireless sensor network.
\newblock {\em International Journal of Distributed Sensor Networks} {\bf
  2012}, {\em 8},~962523.

\bibitem[Dec(2016)]{DW1000APS06}
DecaWave Ltd.
\newblock {\em {DW}1000 metrics for estimation of non line of sight operating
  conditions},  2016.
\newblock APS006 Part 3 Application Note, version 1.1.

\bibitem[Leu et~al.(2022)Leu, Camurati, Heinrich, Roeschlin, Anliker, Hollick,
  Capkun, and Classen]{leu2022ghost}
Leu, P.; Camurati, G.; Heinrich, A.; Roeschlin, M.; Anliker, C.; Hollick, M.;
  Capkun, S.; Classen, J.
\newblock Ghost Peak: Practical Distance Reduction Attacks Against {HRP} {UWB}
  Ranging.
\newblock In Proceedings of the 31st USENIX Sec. Symp. (USENIX Security 22),
  2022, pp. 1343--1359.

\bibitem[Peterseil et~al.(2023)Peterseil, Etzlinger, Khanzadeh, and
  Springer]{peterseil2023}
Peterseil, P.; Etzlinger, B.; Khanzadeh, R.; Springer, A.
\newblock Trustworthiness Score for UWB Indoor Localization.
\newblock In Proceedings of the GLOBECOM 2023 - 2023 IEEE Global Communications
  Conference,  2023, pp. 189--194.
\newblock {\url{https://doi.org/10.1109/GLOBECOM54140.2023.10437828}}.

\bibitem[Poturalski et~al.(2010)Poturalski, Flury, Papadimitratos, Hubaux, and
  Le~Boudec]{poturalski2010cicada}
Poturalski, M.; Flury, M.; Papadimitratos, P.; Hubaux, J.P.; Le~Boudec, J.Y.
\newblock The cicada attack: degradation and denial of service in IR ranging.
\newblock In Proceedings of the Proc. Int. Conf. Ultra-Wideband. IEEE,  2010,
  Vol.~2, pp. 1--4.

\bibitem[Singh et~al.(2021)Singh, Roeschlin, Zalzala, Leu, and
  {\v{C}}apkun]{singh2021security}
Singh, M.; Roeschlin, M.; Zalzala, E.; Leu, P.; {\v{C}}apkun, S.
\newblock Security analysis of {IEEE} 802.15. 4z/{HRP} {UWB} time-of-flight
  distance measurement.
\newblock In Proceedings of the Proc. Conf. Sec. Privacy in Wireless and Mobile
  Netw. ACM,  2021, pp. 227--237.

\bibitem[Xing et~al.(2014)Xing, He, Pecht, and Tsui]{XING2014106}
Xing, Y.; He, W.; Pecht, M.; Tsui, K.L.
\newblock State of charge estimation of lithium-ion batteries using the
  open-circuit voltage at various ambient temperatures.
\newblock {\em Applied Energy} {\bf 2014}, {\em 113},~106--115.
\newblock
  {\url{https://doi.org/https://doi.org/10.1016/j.apenergy.2013.07.008}}.

\bibitem[Zheng et~al.(2016)Zheng, Xing, Jiang, Sun, Kim, and
  Pecht]{ZHENG2016513}
Zheng, F.; Xing, Y.; Jiang, J.; Sun, B.; Kim, J.; Pecht, M.
\newblock Influence of different open circuit voltage tests on state of charge
  online estimation for lithium-ion batteries.
\newblock {\em Applied Energy} {\bf 2016}, {\em 183},~513--525.
\newblock
  {\url{https://doi.org/https://doi.org/10.1016/j.apenergy.2016.09.010}}.

\bibitem[Peterseil et~al.(2022)Peterseil, Märzinger, and
  Etzlinger]{ppeterseil2021}
Peterseil, P.; Märzinger, D.; Etzlinger, B.
\newblock {{UWB} weak-{NLOS} structured dataset}.
\newblock \url{https://github.com/ppeterseil/UWB-weak-NLOS-structured-dataset},
   2022.

\bibitem[Isik et~al.(2020)Isik, Hong, Petrunin, and Tsourdos]{robotics9030066}
Isik, O.K.; Hong, J.; Petrunin, I.; Tsourdos, A.
\newblock Integrity Analysis for GPS-Based Navigation of UAVs in Urban
  Environment.
\newblock {\em Robotics} {\bf 2020}, {\em 9}.
\newblock {\url{https://doi.org/10.3390/robotics9030066}}.

\end{thebibliography}
